\documentclass[usegraphicx,usenatbib,useAMS]{mn2e} 

\usepackage{mathabx}

\newcommand\eqn[1]     {Eq.\,(\ref{#1})}
\newcommand\eqns[2]    {Eqs\,(\ref{#1}) and~(\ref{#2})}

\newcommand\eqnsss[3]  {Eqs\,~(\ref{#1}), ~(\ref{#2}), ~(\ref{#3})}

\newcommand\nn         {\nonumber}

\def\mnras{{Mon.~ Not.~ R.~ Astron.~ Soc.~}}
\def\aj{{Astronomical Journal}}

\def\prd{{Phys.~ Rev.~ D.~}}

\def\apj{{Astrophys.~ J.~}}
\def\apjs{{Astrophys.~ J.~ Suppl.~}}
\def\apjl{{Astrophys.~ J.~ Lett.~}}

\def\pasj{{Publications of the Astronomical Society of Japan}}

\def\mnras{{MNRAS}}

\def\prd{{PRD}}

\def\apj{{ApJ}}
\def\apjs{{ApJS}}
\def\apjl{{ApJL}}
\def\aap{{A\&A}}

\newcommand{\satellitename}[1]{\ensuremath{\mathit{#1}}}
\newcommand{\telescopename}[1]{\ensuremath{\mathrm{#1}}}

\newcommand{\be}{\begin{equation}}
\newcommand{\ee}{\end{equation}}
\newcommand{\ba}{\begin{eqnarray}}
\newcommand{\ea}{\end{eqnarray}}

\newcommand{\mx}{\mbox}
\newcommand{\bm}{\boldmath }
\newcommand{\euclid}{\satellitename{Euclid}}
\newcommand{\lsst}{\telescopename{LSST}}

\newcommand{\planck}{\satellitename{Planck}}
\newcommand{\arcmint}{\mathrm{arcmin}}
\newcommand{\zmed}{z_{\mathrm{med}}}
\newcommand{\thetapix}{\theta_{\mathrm{pix}}}
\newcommand{\arcsect}{\mathrm{arcsec}}
\newcommand{\C}{\mbox{\boldmath $\rm C$}}
\newcommand{\rr}{\mbox{\boldmath $r$}}
\newcommand{\p}{\mbox{\boldmath $p$}}
\newcommand{\m}{\mbox{\boldmath $m$}}
\newcommand{\map}{M_{\rm ap}}
\newcommand{\dd}{\mathrm{d}}

\def\SN{{\mathcal S}/{\mathcal N}}
\def\Fish{{\mathcal F}}

\def\btheta{{\mx {\bm $\theta$}}}
\def\km{{\kappa_{\rm m}}}
\def\kmsq{{\kappa^2_{\rm m}}}
\def\bkm{{\bar\kappa_{\rm m}}}
\def\bkmsq{{\bar\kappa^2_{\rm m}}}
\def\gm{{\gamma_{\rm m}}}
\def\gmsq{{\gamma^2_{\rm m}}}
\def\Mm{{M_{\rm m}}}
\def\Mmi{{M^{i}_{\rm m}}}
\def\Mmii{{M^{i+1}_{\rm m}}}
\def\zm{{z_{\rm m}}}

\def\nn{{\nonumber}}
\def\om{{\Omega_{\rm m}}}
\def\s8{{\sigma_8}}
\def\ns{{n_{\rm s}}}

\def\sss{\scriptscriptstyle}
\def\ts{\textstyle}

\def\deg2{\rm deg^2}
\def\arcmin2{\rm arcmin^2}

\def\rhob{\bar{\rho}}
\def\Hub{{\rm km}s^{-1}{\rm Mpc}^{-1}}
\def\Msol{h^{-1}M_{\odot}}
\def\kpc{\, h^{-1}{\rm kpc}}
\def\Mpc{\, h^{-1}{\rm Mpc}}

\def\Gpccube{\, h^{-3} \, {\rm Gpc}^3}

\def\nbar{\bar{n}}
\def\shn{\sigma_{\gamma}^2/\bar{n}}

\title[Optimized detection of shear peaks in weak lensing maps]
{Optimized detection of shear peaks in weak lensing maps}

\author[Marian, Smith, Hilbert \& Schneider] { Laura
  Marian,$^{1}$\thanks{lmarian@astro.uni-bonn.de}
  Robert~E.~Smith,$^{2,1}$ Stefan Hilbert$^{3,1}$ and Peter
  Schneider$^{1}$ \\ $^1$ Argelander-Institute for Astronomy, Auf dem
  H\"ugel 71, D-53121 Bonn, Germany\\ $^2$ Institute for Theoretical
  Physics, University of Zurich, Zurich CH 8037\\ $^3$ Kavli Institute
  of Particle Astrophysics and Cosmology (KIPAC), Stanford University,
  452 Lomita Mall, Stanford, CA 94305, \\ and SLAC National
  Accelerator Laboratory, 2575 Sand Hill Road, M/S 29, Menlo Park, CA
  94025 }

\begin{document}

\maketitle


\begin{abstract}
We present a new method to extract cosmological constraints from weak
lensing (WL) peak counts, which we denote as `the hierarchical
algorithm'. The idea of this method is to combine information from WL
maps sequentially smoothed with a series of filters of different size,
from the largest down to the smallest, thus increasing the
cosmological sensitivity of the resulting peak function. We compare
the cosmological constraints resulting from the peak abundance
measured in this way and the abundance obtained by using a filter of
fixed size, which is the standard practice in WL peak studies. For
this purpose, we employ a large set of WL maps generated by
ray-tracing through $N$-body simulations, and the Fisher matrix
formalism. We find that if low-$\SN$ peaks are included in the
analysis ($\SN\sim 3$), the hierarchical method yields constraints
significantly better than the single-sized filtering. For a large
future survey such as $\euclid$ or $\lsst$, combined with information
from a CMB experiment like $\planck$, the results for the hierarchical
(single-sized) method are: {$\Delta \ns=0.0039\, (0.004);\, \Delta
  \om=0.002\, (0.0045);\, \Delta \s8=0.003\, (0.006);\, \Delta
  w=0.019\, (0.0525)$}.  This forecast is conservative, as we assume
no knowledge of the redshifts of the lenses, and consider a single
broad bin for the redshifts of the sources. If only peaks with
$\SN\geq 6$ are considered, then there is little difference between
the results of the two methods. We also examine the statistical
properties of the hierarchical peak function: Its covariance matrix
has off-diagonal terms for bins with $\SN \leq 6$ and aperture mass of
$M<3\times10^{14}\, \Msol$, the higher bins being largely uncorrelated
and therefore well described by a Poisson distribution.
\end{abstract}



\section{Introduction}
\label{I}
For more than a decade, weak gravitational lensing (WL) has been
considered a powerful probe for testing cosmology due to its potential
to map the 3D matter distribution of the Universe in an unbiased way,
independent of baryonic matter tracers.

Several surveys have already demonstrated the ability of WL to
constrain the cosmological model through cosmic shear measurements,
e.g.  the Cerro Tololo Inter-American Observatory (CTIO) lensing
survey \citep{Jarvisetal2003, Jarvisetal2006}, the Garching-Bonn deep
survey \citep[GaBoDS,][]{Hetterscheidtetal2007}, and the
Canada-France-Hawaii Telescope Legacy Survey
\citep[CFHTLS,][]{Hoekstraetal2006, Sembolonietal2006}.

Among the WL probes are shear peaks, regions of high signal-to-noise
($\SN$) in shear maps that can be produced by individual clusters or
by the alignment of several smaller objects on the line of sight. The
abundance of shear peaks is as sensitive to cosmology as the cluster
mass function \citep{MarianSB2009, MarianSB2010, Kratochviletal2010,
  Kratochviletal2011, Yangetal2011}. Clusters are one of the four
most promising tools to measure dark energy \citep{DETF}, together
with supernova surveys, baryonic acoustic oscillations, and WL
surveys. Therefore, the WL peak function is equally promising in
principle. There are two major advantages of WL peaks over clusters:
(i) WL peaks will come for free with any future lensing survey; (ii)
one can very reliably calibrate the abundance of peaks with cold dark
matter (CDM) simulations and proceed with direct comparisons to data
measurements, thus bypassing the thorny issue of the mass-observable
relation. Shear peak signals need not be translated into virial masses
in order to be able to extract cosmological information from their
abundance \citep{DietrichHartlap2010}. A disadvantage is the absence
of an analytical framework for the WL peaks \citep[though see the
  recent work of][]{Maturietal2010}.

Detections of shear peaks in WL data are exemplified in the works of
\citet{Dahle2006, Schirmeretal2007, Berge2008, Abateetal2009}. However,
the exploration of the shear signal of clusters has been focused
mostly on mass determination, as for instance in the recent work of
\citet{Okabe2010} and \citet{Israeletal2010}.

Since the introduction of the aperture mass by \cite{Schneider1996},
there have been many studies of filters optimal for peak detection and
of the impact of large-scale structure (LSS) projections on cluster
mass reconstructions \citep[e.g.][]{Metzleretal2001, Hoekstra2001,
  Hamanaetal2004, Cloweetal2004, TangFan2005, Maturietal2005,
  HennawiSpergel2005, MarianSB2010, BeckerKravtsov2011,
  Gruenetal2011}. The general agreement is that cluster masses derived
from WL measurements are affected by both correlated and uncorrelated
LSS projections, as well as by departures of the density profiles of
real clusters from the assumed spherical models. These effects cause
scatter and bias in the predicted and measured $\SN$ of the
clusters. Nonetheless WL mass reconstructions retain the attractive
feature of being able to rely on numerical simulations for accurate
predictions of such biases.

In this paper we address a more general question related to the
abundance of shear peaks. Given the upcoming WL surveys such as the
Kilo-Degree Survey \citep[KiDS, ][]{Kuijken2010}, the Dark Energy Survey
\citep[DES,][]{DES2005}, the Large Synoptic Survey Telescope (LSST)
survey \citep[][]{lsst2009}, or the $\euclid$ survey
\citep[][]{Euclid2011}, it will be possible to
measure the shear peak function: what is the optimal way to do this?

The standard approach to peak detection is: A given shear map is
smoothed with a given filter function; in the smoothed map, one looks
for points of local maximum which have $\SN$ higher than a certain
chosen threshold value, and one selects these points as `peaks'. This
procedure is dependent on the filter used. There have been many
studies on the shape of filters that maximize the $\SN$ assuming
certain shapes for the peak signal and various types of noise such as
shape noise or projection noise \citep[e.g.][]{HennawiSpergel2005,
  Maturietal2005, Gruenetal2011}.
  
Less attention has been directed towards the size of filters. Indeed
in most studies, the peak abundances are measured using a single-sized
filter \citep[e.g.][]{Hamanaetal2004, DietrichHartlap2010}, though it
is clear that each size will lead to a different peak function. For a
different approach using wavelets, see \cite{Piresetal2009}. Here we
propose a method that we call `hierarchical algorithm': A shear map is
smoothed with several filters of the same shape but different size,
from the largest to the smallest. We show that by taking into account
the extended information from such multi-scale filtering, one can
assign in the context of an assumed halo paradigm, e.g. the
\cite{NFW1997} model, a unique $\SN$ and (redshift-dependent) mass to
the detected peaks. We use the Fisher matrix formalism and a large set
of simulated WL maps to show that the cosmological constraints derived
from the hierarchical peak function are much improved compared to
those obtained using a filter of the same shape but only one size.

The paper is structured as follows. In section \S\ref{II}, we present
the $N$-body simulations and the ray-tracing performed to
generate the WL maps employed in this study. In \S\ref{III} we explain
the hierarchical scheme and the filter that we adopt. The results of
this work are presented in \S\ref{IV}, along with a WL-peaks Fisher
forecast for surveys like \euclid\ and \lsst, the first to be
obtained from simulation measurements. In \S\ref{V} we summarize and
conclude.

\section{Numerical simulations and ray-tracing}
\label{II}
\begin{table*}
\caption{{\tt zHORIZON} cosmological parameters. Columns
are: density parameters for matter, dark energy and baryons; the
equation of state parameter for the dark energy;
normalization and primordial spectral index of the power spectrum;
dimensionless Hubble parameter.
\label{tab:zHORIZONcospar}}
\vspace{0.2cm}
\centering{
\begin{tabular}{c|ccccccc}
\hline 
Cosmological parameters & $\om$ & $\Omega_{\rm DE}$ & $\Omega_b$ & $w$  &  
$\s8$  & $n$ &  $H_{0} [\Hub]$ \\
\hline
{\tt zHORIZON-I      }  & 0.25\ &  0.75 & 0.04 &  -1  &  0.8  & 1.0 & 70.0\\
{\tt zHORIZON-V1a/V1b}  & 0.25\ &  0.75 & 0.04 &  -1  &  0.8  & 0.95/1.05 & 70.0\\
{\tt zHORIZON-V2a/V2b}  & 0.25\ &  0.75 & 0.04 &  -1  &  0.7/0.9  & 1.0 & 70.0\\
{\tt zHORIZON-V3a/V3b}  & 0.2/0.3\ &  0.8/0.7 & 0.04 &  -1  &  0.8  & 1.0 & 70.0\\
{\tt zHORIZON-V4a/V4b}  & 0.25\ &  0.75 & 0.04 &  -1.2/-0.8  &  0.8  & 1.0 & 70.0\\
\end{tabular}}
\end{table*}
\begin{table*}
\caption{\small {\tt zHORIZON} numerical parameters. Columns are: number of
particles, box size, particle mass, force softening, number of realizations, 
and total simulated  volume. 
\label{tab:zHORIZONsimpar}}
\vspace{0.2cm}
\centering{
\begin{tabular}{c|cccccc}
\hline
Simulation Parameters & 
$N_{\rm part}$ & $L_{\rm sim}\,[\Mpc]$ & $m_p [\Msol]$ &
 {$l_{\rm soft}\,[\kpc]$} & $N_{\rm ensemb}$ & $V_{\rm tot}[\Gpccube]$\\
\hline
{\tt zHORIZON-I}  & $750^3$ & 1500 & $5.55\times 10^{11}$ & 60 & 8 & 27 \\
{\tt zHORIZON-V1, -V2, -V4} & $750^3$ & 1500 & $5.55\times 10^{11}$ & 60 & 4  &  13.5 \\
{\tt zHORIZON-V3a} & $750^3$ & 1500 & $4.44\times 10^{11}$ & 60 & 4  &  13.5 \\
{\tt zHORIZON-V3b} & $750^3$ & 1500 & $6.66\times 10^{11}$ & 60 & 4  &  13.5\\
\end{tabular}}
\end{table*}

We generated WL maps from ray-tracing through $N$-body simulations. We
used 8 simulations which are part of a larger suite performed on the
zBOX-2 and \mbox{zBOX-3} supercomputers at the University of
Z\"{u}rich. For all realizations 11 snapshots were output between
redshifts $z=[0,2]$; further snapshots were at redshifts
$z=\{3,4,5\}$. We shall refer to these simulations as the {\tt
  zHORIZON} simulations, and they were described in detail in
\cite{Smith2009}.

Each of the {\tt zHORIZON} simulations was performed using the
publicly available {\tt Gadget-2} code \citep{Springel2005}, and
followed the nonlinear evolution under gravity of $N=750^3$ equal-mass
particles in a comoving cube of length $L_{\rm sim}=1500\Mpc$; the
softening length was $l_{\rm soft}=60\, \kpc$. The cosmological model
was similar to that determined by the WMAP experiment
\citep{Komatsuetal2009short}. We refer to this cosmology as the fiducial
model. The transfer function for the simulations was generated using
the publicly available {\tt cmbfast} code
\citep{SeljakZaldarriaga1996}, with high sampling of the spatial
frequencies on large scales. Initial conditions were set at redshift
$z=50$ using the serial version of the publicly available {\tt 2LPT}
code
\citep{Scoccimarro1998,Crocceetal2006}. Table~\ref{tab:zHORIZONcospar}
summarizes the cosmological parameters that we simulated and
Table~\ref{tab:zHORIZONsimpar} summarizes the numerical parameters
used.

For the Fisher matrix study of peak counts, we employed another series
of simulations. Each of the new set was identical in every way to the
fiducial model, except that we have varied one of the cosmological
parameters by a small amount. For each new set we have generated 4
simulations, matching the random realization of the initial Gaussian
field with the corresponding one from the fiducial model. The four
parameter variations were: $\{n\rightarrow (0.95, 1.05
),\,\s8\rightarrow (0.7, 0.9),\,\om\rightarrow (0.2, 0.3),\,
w\rightarrow (-1.2, -0.8) \}$, and we refer to each of the sets as
{\tt zHORIZON-V1a,b},\dots,{\tt zHORIZON-V4a,b}, respectively. The
 details are summarized in Tables \ref{tab:zHORIZONcospar} and
\ref{tab:zHORIZONsimpar}.

For the WL simulations, we considered a survey similar to \euclid{}
\citep{Euclid2011} and to \lsst{} \citep{lsst2009}, with: an
rms $\sigma_{\gamma}=0.3$ for the intrinsic image ellipticity, a
source number density $\nbar =40\,\arcmint^{-2}$, and a redshift
distribution of source galaxies given by: 
\be
{\mathcal P}(z)={\cal N}(z_0, \beta)\,z^2\exp[-(z/z_0)^{\beta}], 
\label{eq:source_distribution}
\ee
where the normalization constant ${\cal N}$ insures that the integral
of the source distribution over the source redshift interval is
unity. If this interval extended to infinity, then the normalization
could be written analytically as: ${\cal
  N}=3/(z_0^3\,\Gamma[(3+\beta)/\beta])$. There is a small difference
between this value and what we actually used, due to the fact that we
considered a source interval of $[0, 3]$. We took $\beta=1.5$, and
required that the median redshift of this distribution be $\zmed=0.9$,
which fixed $z_0\approx 0.64$, and gave a mean of $z_{\rm mean}=0.95$.

From each $N$-body simulation we generated 16 independent fields of
view. Each field had an area of $12 \times 12\,\deg2$ and was tiled by
$4096^2$ pixels, yielding an angular resolution $\thetapix
=10\,\arcsect$. For each variational model, the total area was of
$\approx 9000\, \deg2$, while for the fiducial model it was of
$\approx 18000\, \deg2$. The effective convergence $\kappa$ in each
pixel was calculated by tracing a light ray back through the
simulation with a multiple-lens-plane ray-tracing algorithm
\citep{Hilbertetal2007, Hilbertetal2009}. Gaussian shape noise with
variance $\sigma_{\gamma}^2 /(\nbar\, \thetapix^{2})$ was then added
to each pixel, creating a realistic noise level and correlation in the
filtered convergence field \citep{HilbertMetcalfWhite2007}. We keep
the shape noise configuration fixed for each field in different
cosmologies, in order to minimize its impact on the comparisons of the
peak abundances measured for each cosmology.

\section{Smoothing weak lensing maps}
\label{III}
\subsection{ A matched filter }
\label{III1}
To find peaks in WL maps, we smooth the latter with an aperture-mass
filter \citep{Schneider1996, Schneideretal1998}. The smoothed
convergence map is a convolution between the filter function and the
$\kappa/\gamma$ field of our simulations:
\be 
\map(\btheta_0)=\int \dd^2\theta\,
U(\btheta_0-\btheta)\,\kappa(\btheta)=\int \dd^2\theta\,
Q(\btheta_0-\btheta)\,\gamma(\btheta),\hspace{0.2cm}
\label{eq:smoothed_map}
\ee
where $\kappa$ is the convergence, $\gamma$ is the tangential shear
field, and $U$ and $Q$ are aperture filters for convergence and shear
respectively. $\btheta_0$ is an arbitrary point. Aperture mass filters
are compensated, which for a spherically symmetric function can be
expressed through the equation:
\be
\int_{0}^{\theta_{\rm A}} \dd\theta \,\theta \,U(\theta)=0,
\ee
where $\theta_{\rm A}$ is the compensation radius. In the presence of
the ellipticity noise of the source galaxies, it can be shown that an
optimal and compensated filter is given by
\be
U(\theta)=\mathcal{C} \, \frac{\km(\theta)-\bkm(\theta_{\rm A})}{\shn},
\label{eq:filterC_def}
\ee
where $\mathcal{C}$ is an arbitrary normalization constant, $\km$ is
the adopted model for the convergence profile of peaks, i.e. NFW or
similar, and $\shn$ is the shape noise variance per ellipticity
component. The mean convergence inside a radius $\theta$ is defined by
$\bar\kappa(\theta)=2/\theta^2\,\int_0^{\theta} d\theta'\,\theta'
\kappa(\theta')$. The analogue filter function for the shear field is
given by
\be
Q(\theta)=\mathcal{C} \, \frac{\gm(\theta)}{\shn},
\label{eq:filterG_def}
\ee
where $\gm$ is the assumed tangential shear model of the peaks. Under
the assumption that the shape noise is the dominant source of noise in
the measurements, this filter is optimal because it maximizes the
signal-to-noise ($\SN$) at the location of a peak with the
convergence/shear profile $\km$/$\gm$. From
\eqns{eq:smoothed_map}{eq:filterC_def}, the $\SN$ can be written as
\be
\SN(\btheta_0)=\sqrt{\frac{\nbar}{\sigma_{\gamma}^2}}\, \frac{\int d^2\theta\,
  [\km(\theta)-\bkm(\theta_{\rm A})]\kappa(\btheta_0-\btheta)} 
     {\sqrt{\int d^2\theta\,[\km(\theta)-\bkm(\theta_{\rm A})]^2}},
\label{eq:SNc}
\ee
or in terms of the shear
\be
\SN(\btheta_0)=\sqrt{\frac{\nbar}{\sigma_{\gamma}^2}}\, \frac{\int d^2\theta\,
\gm(\theta)\,\gamma(\btheta_0-\btheta)}{\sqrt{\int d^2\theta 
    \, \gmsq(\theta)}}.
\label{eq:SNs}
\ee
Note that the $\SN$ {\it does not} depend on the arbitrary
normalization constant $\cal C$. 

Although we do not make a comparison between peaks and clusters, and
indeed do not use any information on the simulation halos in this
study, our choice of $\cal C$ provides insight into the halos that
generate the peaks (though of course not all the peaks will correspond
to a halo). If $\btheta_0$ denotes the location of a peak formed by a
halo of mass $\Mm$, redshift $\zm$, and profile
$\kappa(\theta)=\km(\theta ; \Mm, \zm)$, then we require that the
amplitude of the smoothed map at the location of this peak be exactly
$\Mm$: $\map(\btheta_0)=\Mm$. In this case, $\cal C$ is given by
\citep{MarianBernstein2006}:
\be
{\cal C}(\Mm) = \shn\,\,\frac{\Mm}{\int d^2\theta \,
  \kmsq(\theta)-\pi\theta^2_{A}\bkmsq(\theta_{A})}.
\label{eq:normC}
\ee
In the above equation, $\km(\theta)=\km(\theta ; \Mm, \zm)$ and it is
assumed that the radial integral has an upper limit of
$\theta_{A}$. The latter applies also to
\eqnsss{eq:SNc}{eq:SNs}{eq:smoothed_map2}. Note that for the shear
filter the normalization is the same, since
\be 
2\pi \int_0^{\theta_{A}} d\theta \,\theta\,
\kmsq(\theta)-\pi\theta^2_{A}\,\bkmsq(\theta_{A})=2\pi \int_0^{\theta_{A}}
d\theta \, \theta\, \gmsq(\theta).
\ee
Our analysis was performed on convergence maps, and so in the
following, we shall focus on the latter. Inserting
\eqns{eq:filterC_def}{eq:normC} into \eqn{eq:smoothed_map}, we write
down the amplitude of the smoothed map for our particular choice of
matched aperture filter:
\ba
\map(\btheta_0)& = &\Mm\frac{\int d^2\theta\,[\km(\theta)-
    \bkm(\theta_{\rm A})]\kappa(\btheta_0-\btheta)}{\int d^2\theta \,
  \kmsq(\theta)-\pi\theta^2_{A}\bkmsq(\theta_{A})}\nn \\
 &=&  \sqrt{{\cal C}(\Mm)\, \Mm}\: \SN (\btheta_0).\hspace{2cm}
\label{eq:smoothed_map2}
\ea
Within the validity bounds of our model, i.e. the peak is indeed
generated by an NFW halo of that mass and redshift,
\eqn{eq:smoothed_map2} represents an unbiased estimator for
mass. 

Finally, we assume a relation between the model mass $\Mm$ and the
compensation radius: We take the latter to be the angular scale
subtended by the virial radius of a halo with mass $\Mm$, redshift
$\zm$, and convergence profile $\km$: \be \theta_A=R_{\rm vir}(\Mm,
\zm)/D_{A}(\zm), \ee where $D_{A}(\zm)$ is the angular diameter
distance to $\zm$. This choice enables us to connect the `size' of the
filter, i.e. the aperture radius, with the `mass' of the filter $\Mm$,
using the standard relation between mass and virial radius provided by
models of structure formation, such as NFW for instance. Given the
source distribution in \eqn{eq:source_distribution}, we take
$\zm=0.3$. This is just a clarification of what we mean by size and
mass of the filter, all the necessary details will be provided in
\S\ref{III3}.
\begin{figure}
\centering
\includegraphics[scale=0.45]{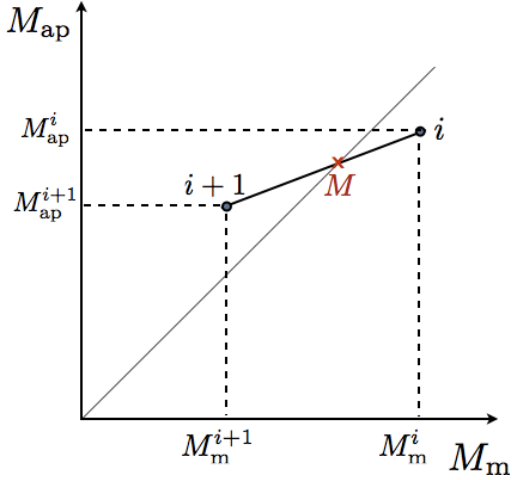}
\caption {Matching between the filter and peak profiles in the
  hierarchical method. The larger filter $\Mmi$ yields an aperture
  mass $\map^i<\Mmi$ while the next filter in the sequence, of smaller
  size $\Mmii$, gives an aperture mass $\map^{i+1}>\Mmii$. The
  solution $M$ to \eqn{eq:match} is found through interpolation to be
  \eqn{eq:Mtrue}. We take this to be the `true' mass of the peak. }
\label{fig:cartoon}
\end{figure}
\subsection{ Hierarchical Filtering }
\label{III2}
We shall now describe our method to detect shear peaks and assign them
masses and $\SN$. This was already implemented in our previous works
\cite{MarianSB2009, MarianSB2010, Marianetal2011}.

As mentioned in \S\ref{III1}, smoothed maps are obtained by convolving
the convergence field with a filter, e.g. \eqn{eq:smoothed_map}. Peaks
are detected as local maxima in the smoothed maps, i.e. points with
amplitude higher than that of their 8 neighbors, where the amplitude
is given by \eqn{eq:smoothed_map}, or in our particular case, by
\eqn{eq:smoothed_map2}. Medium or large peaks will still be local
maxima even when smoothed with filters of size much larger or smaller
than the peak radius. But the $\SN$ and amplitude associated to the
peak will be quite different for a range of filter sizes spanning 1-2
orders of magnitude. Therefore, there is some degree of arbitrariness
when trying to classify the abundance of WL peaks in terms of their
$\SN$ or amplitude: the answer will depend on the filter size. Also,
if the peaks are small, then a large filter may render them quite
indistinguishable from spurious shape-noise peaks. The strengths of
our top-down approach are:
\begin{enumerate}
 \item It uses
  filters of several sizes, which will increase the scale range of the
  detected peaks and therefore the cosmological information of the
  peak counts. 
  \item It uses an interpolation scheme for the results of the
    smoothing with each filter to accurately establish a unique value
    for the $\SN$ and amplitude $\map$ of each peak. Thus the
    ambiguity of classifying peaks in terms of their $\SN$ is removed,
    and one obtains a `general' peak function, as opposed to a
    different peak function for each filter size employed in
    smoothing.
\end{enumerate}

The concrete steps that we take are as follows. We smooth the maps
with a sequence of filters of different sizes (masses), in a
hierarchical fashion, from the largest to the smallest size. The
purpose is to determine the `mass' of the peaks, i.e. the filter size
which matches best the size of the peaks:
\be
\map(\btheta_0)=\Mm \,,
\label{eq:match}
\ee
where $\btheta_0$ denotes the location of a detected peak. For each
filter $i$ in the sequence, the peaks are selected so that: (a)
$\map^i \geq \Mmi$; (b) $\SN^i \geq (\SN)_{\rm min}$, where $\Mmi$ is
the size of the filter. $\map^i$ and $\SN^i$ are the aperture mass and
$\SN$ defined in \eqns{eq:SNc}{eq:smoothed_map2} corresponding to this
particular filter, and we choose $(\SN)_{\rm min}=3$ as a detection
threshold.

Equation~(\ref{eq:match}) is not likely to be satisfied by any
particular filter in the sequence, hence we find its solution by
interpolating between the results of smoothing with different filters
in the sequence. This is illustrated in Figure \ref{fig:cartoon}.
Suppose there is a peak of true mass $M$ and $\SN$ (true according to
the assumed model). Then there will be two consecutive filters in the
sequence, $i$ and $i+1$, for which the following relations are true:
$$
\Mmi>\map^{i}>M>\map^{i+1}>\Mmii,\hspace{0.2cm}
\SN^i>\SN>\SN^{i+1}.\hspace{0.1cm}
$$
Since the aperture mass of the peak obtained from consecutive filters
varies gently with the size of the filter, we can use linear
interpolation to write down the solution for the true mass, i.e. the
solution of \eqn{eq:match}.
\be
M=\frac{\Mmi \map^{i+1}-\Mmii \map^{i}}{\Mmi - \Mmii -\map^{i} + \map^{i+1}}.
\label{eq:Mtrue}
\ee
Once the true mass is determined, we use
\eqns{eq:smoothed_map2}{eq:match} to determine the $\SN$ of the peak:
\be
\SN=\sqrt{M/{\cal C}(M)}
\label{eq:SNtrue}
\ee
Given our two selection criteria, in the above example the point of
local maximum will be selected as a peak by the filter $i+1$, but not
by its predecessor $i$, if the signal to noise will also be above the
detection threshold. It will also appear as a peak in the maps
smoothed with filters $<\Mmii$, provided that the same $\SN$
requirement is fulfilled. In order to be able to carry out the
interpolation scheme, for each filter used we record the aperture
mass, $\SN$, and the 2D location of the peaks. A peak might slightly
change its coordinates in maps smoothed with different-sized filters;
we take this into account, and allow for variations of up to 4 pixels
in the ${\bf\hat x,\,\hat y}$ directions of the map (a pixel has
$\thetapix =10\,\arcsect$). We also record those points of maximum
where the aperture mass is smaller, but not much smaller than the mass
of the respective filter; to be specific, the points obeying the
condition: $0.6\Mmi<\map^i<\Mmi$. These points we call `pseudopeaks'
and they are likely to be selected as peaks by the next filter in the
sequence; therefore, they are useful for the interpolation that we
perform later. The value $0.6$ is of no particular significance, it is
suitable for the logarithmically-spaced sequence of filters that we
apply, based on several trials.

Finally, the processing of the peaks resulting from the smoothing with
the hierarchical sequence of filters, consists of the following steps:
1. We exclude from the maps those peaks already selected by a larger
filter; 2. We apply the interpolation scheme to assign the remaining
peaks a unique aperture mass, according to \eqn{eq:Mtrue}. We then use
this mass to compute the $\SN$ value, according to \eqn{eq:SNtrue};
3. We exclude those peaks that are within the virial radius of a
larger peak, since in many of such cases, the second peak is just an
artifact of the smoothing or we simply deal with a very clumpy halo
that is split by the smoothing into a large peak and some small
ones. We thus remove the problem of `peaks-in-peaks' and also do not
count substructures as independent halos. This is done for the purpose
of obtaining a `clean' peak function, but ultimately such events
concern only small peaks and we have checked that the cosmological
constraints derived from the counts are not significantly altered by
these exclusions.
\begin{figure*}
\centering
\includegraphics[scale=0.9]{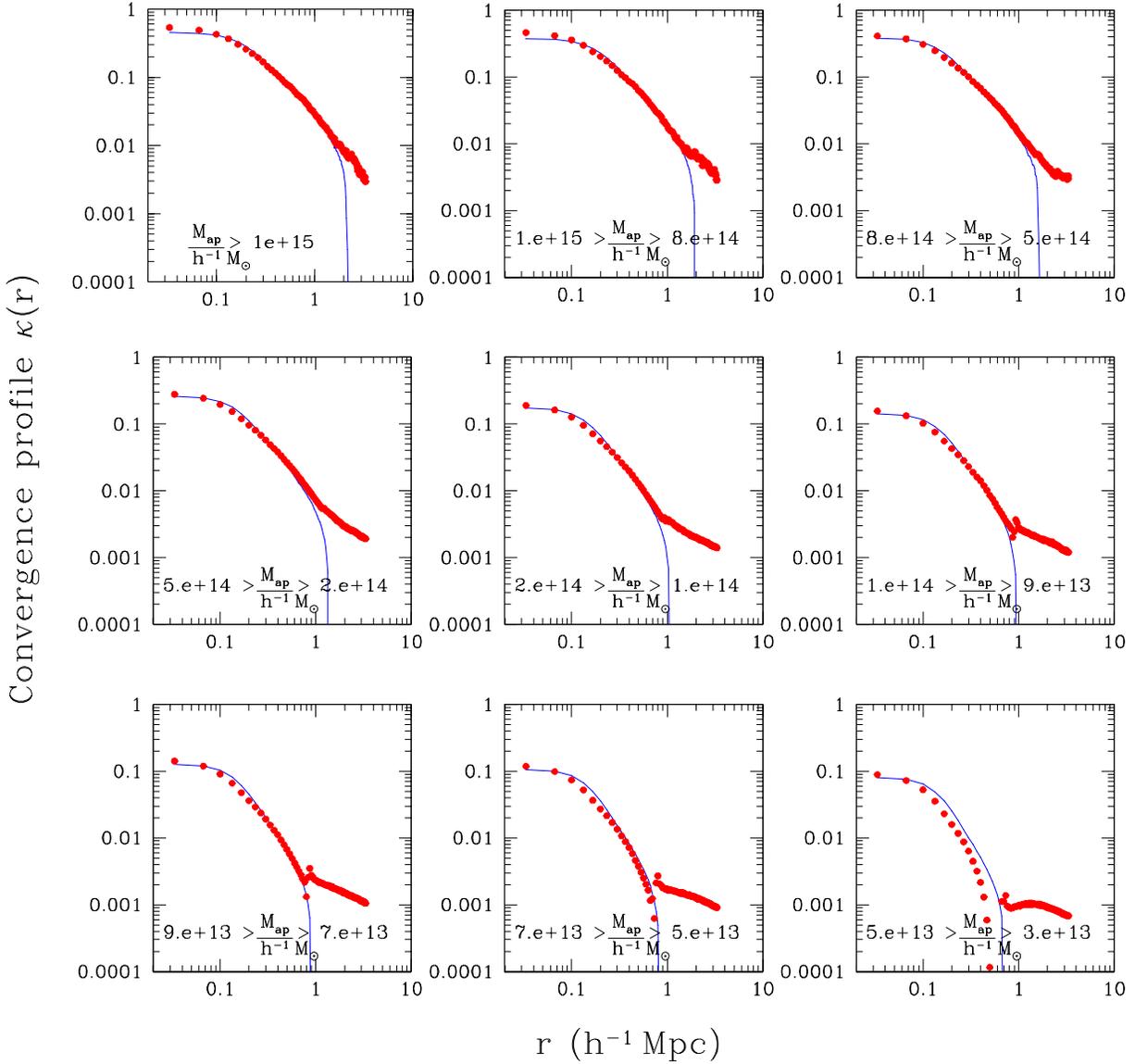}
\caption{Comparison between measured and theoretical convergence
  profiles for the unsmoothed fiducial cosmology maps, with the
  sources at redshift 1. The red points represent the average of the
  measured profiles around the centres of the peaks detected with the
  hierarchical method of \S\ref{III2}. The peaks have been binned
  according to the mass assigned through that method, and the panels
  depict different mass bins. The blue solid lines represent the
  theoretical profile from \eqn{eq:kmodel2}, for the mean mass of the
  bin. The redshift of the filter is set at 0.3, and the error bars
  are on the mean of 128 fields. The maps are free of shape noise.}
\label{fig:profiles}
\end{figure*}
\subsection{ Model specifications }
\label{III3}
We base our halo model on the NFW density profile:
\be
\rho_{\rm \sss NFW}(r)=\rhob\,\delta_c\,\left[\frac{r}{r_s}
  \left(1+\frac{r}{r_s}\right)^2\right]^{-1},
\label{eq:rho3D}
\ee 
where $\rhob$ is the mean matter density of the Universe, $\delta_c$
is the characteristic overdensity, and $r_s$ is the scale radius. We
adopt the Sheth-Tormen (ST) definition of mass
\citep{ShethTormen1999}: $M_{\rm vir}=4 \pi R^3_{\rm vir} \Delta_{\rm
  vir} \rhob/3$, i.e. we use the mean matter density to define the
overdensity for halo formation, as opposed to the critical density,
$\rho_{\rm crit}$. The two are related by $\rhob=\Omega_{\rm
  m}\rho_{\rm crit}$. $\Delta_{\rm vir}=200$ for ST and
NFW. Integrating \eqn{eq:rho3D} to obtain the virial mass and using
the above definition for the latter, one arrives at the following
expression for the characteristic overdensity:
\be
\delta_c=\frac{\Delta_{\rm vir}\,c^3/3}{\log(1+c)-c/(1+c)},
\ee
where the concentration parameter is defined by $c=R_{\rm
  vir}/r_s$. In the $\Lambda$CDM model, ST and NFW halos have the same
density profile, but ST halos have larger cut-off radii and
concentration parameters than NFW ones. For the concentration
parameter we employed the numerical prescription of
\cite{Gaoetal2008}, whilst to translate NFW to ST parameters, we used
the approach of \cite{SmithWatts2005}.

We use the truncated convergence profile resulting from this profile,
i.e. we limit the projection of the 3D density along the line of sight
to a region delimited by the virial radius:
\be
\kappa_{\rm \sss NFW}(r_{\perp})=\frac{1}{\Sigma_{\rm crit}}\int_{-\sqrt{R^2_{\rm vir}-r^2_{\perp}}}^
      {\sqrt{R^2_{\rm vir}-r^2_{\perp}}} dz \, \rho_{\rm \sss NFW}(\sqrt{r_{\perp}^2+z^2}),
\ee
with $\Sigma_{\rm crit}$ being the critical surface density for lensing.
The above equation can be rewritten as
\be
\kappa_{\rm \sss NFW}(x)=\frac{\ts2\,r_{s}\,\delta_{c}\,\rhob}{\Sigma_{\rm crit}}\,f(x),
\ee
where $x=r_{\perp}/r_s$ is adimensional, and the function $f$ depends
on cosmology only through the concentration parameter
\citep{Hamanaetal2004}:
\be
f(x)= \left\{ \begin{array}{lcl}
-\frac{\ts{(c^{2}-x^{2})^{1/2}}}{\ts{(1-x^{2})\,(1+c)}} 
+ \frac{\ts{ \cosh^{-1}\left(\frac{x^{2}
+c}{x(1+c)}\right)}}{\ts{(1-x^{2})^{3/2}}},\: x<1 \vspace{0.2cm}\\
\frac{\ts{(c^2-1)^{1/2}}}{\ts{3 (1+c)}}\, \left(1+\frac{\ts{1}}{\ts{1+c}} \right),\: x=1 \vspace{0.2cm}\\
-\frac{\ts{(c^{2}-x^{2})^{1/2}}}{\ts{(1-x^{2})\,(1+c)}} 
- \frac{\ts{\cos^{-1}\left(\frac{x^{2} +c }{x(1+c)}\right)}}{\ts{(x^{2}-1)^{3/2}}},\: x>1 
 \vspace{0.2cm}\\
0, \: x>c.
\end{array}\right.
\label{eq:STkappa}
\ee
The model that we assume for the $\kappa$ of halos is a convolution of
the NFW convergence profile defined by \eqn{eq:STkappa} with a
two-dimensional (2D) Gaussian function with the width of the order of
the softening length of the simulations:
\be
\km(\btheta)=\int d^2\theta'\,\kappa_{\rm \sss NFW}(\btheta')\,G_{\rm \sss 2D}(\btheta-\btheta').
\label{eq:kmodel1}
\ee
The above equation can be rewritten as
\ba
\km(\btheta)=\frac{\exp\left(\frac{-\theta^2}{2\sigma^2_{\rm \sss G}}\right)}
   {\sigma^2_{\rm \sss G}} \times \hspace{4cm} \nn \\
 \int_0^\infty d\theta' \, \theta' \kappa_{\rm \sss NFW}(\theta')\exp\left(\frac{-\theta'^2}
    {2\sigma^2_{\rm \sss G}}\right) I_0\left(\frac{-\theta\, \theta'}
    {\sigma_{\rm \sss G}^2}\right),
\label{eq:kmodel2}
\ea
where $\sigma_{\rm\sss G}$ is the width of the Gaussian function, and
$I_0$ is the modified Bessel function of order $0$. The dependence on
the lens redshift is implicit for both $\kappa_{\rm \sss NFW}$ and
$\sigma_{\rm\sss G}$.
\begin{figure}
\centering
\includegraphics[scale=0.43]{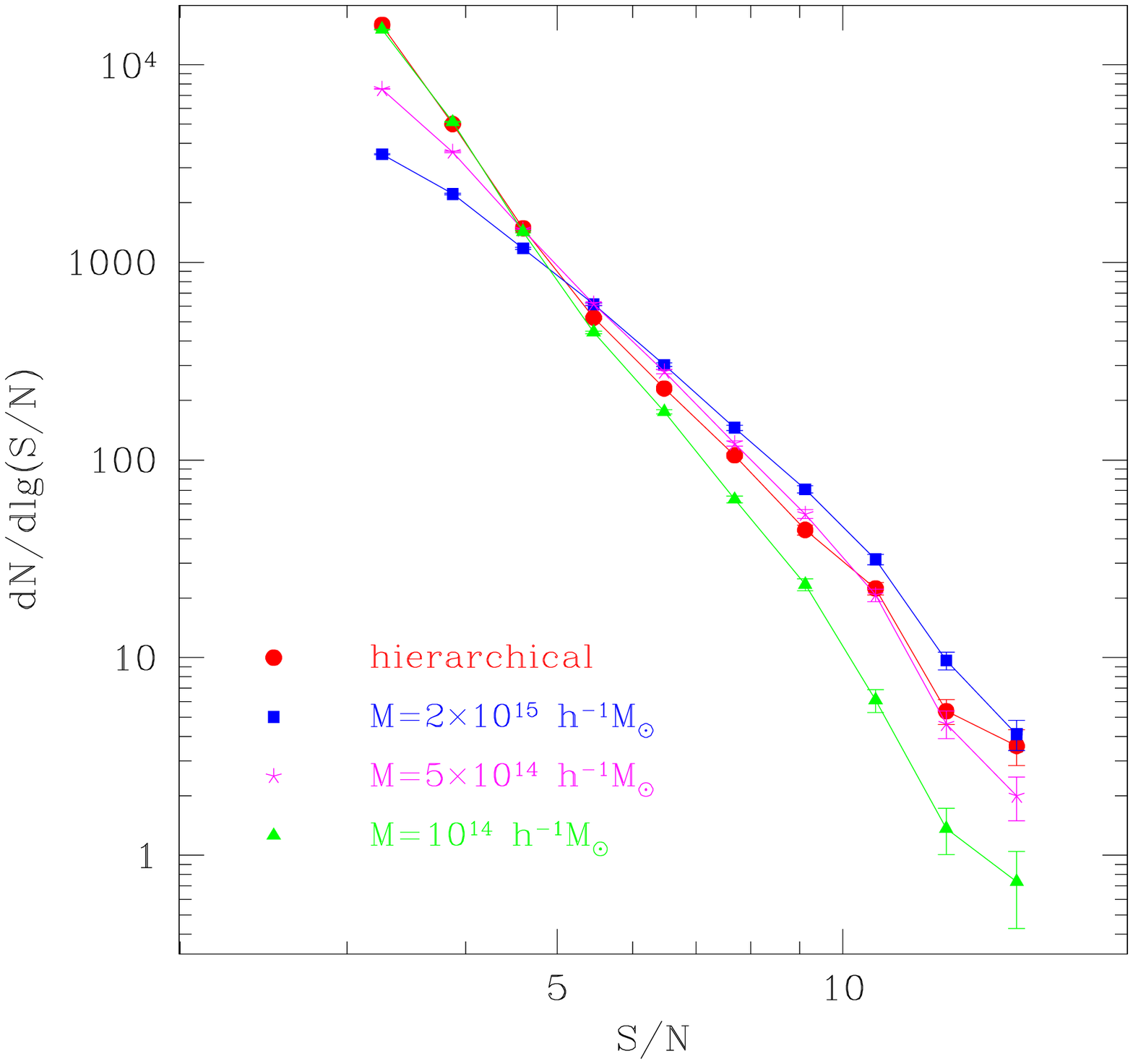}
\caption{The number of peaks per unit $\SN$ for an area of
  $144\,\deg2$. Different symbols and colors denote different filter
  sizes: red circles correspond to the hierarchical filtering, and
  blue squares/purple stars/green triangles to the
  $M={2\times10^{15},\,5\times10^{14},\, 10^{14}}\Msol$ filters
  respectively. The redshift of the filter is kept fixed throughout
  the analysis at $\zm=0.3$. The results are the mean of the functions
  measured from 128 fields of the fiducial cosmology, and the error
  bars are on the mean.}
\label{fig:pfSN}
\end{figure}
This model choice accounts for the finite resolution of the numerical
simulations. The convolution in \eqn{eq:kmodel2} has a similar effect
to `coring' the convergence profile, i.e. making it flat in the centre
of the cluster, where the WL regime breaks down and measurements are
very difficult to obtain. For numerical simulations, cored profile
models are desirable because one cannot resolve structures below the
softening length. Lastly, \eqn{eq:kmodel2} alleviates uncertainties in
the location of the centre of the peak, which could lead to large
discrepancies between measured and theoretical profiles, if the latter
have a cusp at the centre, e.g. like NFW.
Therefore, we take the width of the Gaussian present in the
convolution to be $\sigma_{\rm \sss G}=\alpha \, l_{\rm soft}$, where
$l_{\rm soft}$ is the softening length of the simulations, $\alpha=2$
for ST halos with $M\geq7\times 10^{14} \Msol$, and $\alpha=1.5$
for $M < 7\times 10^{14} \Msol$. For the redshift $\zm=0.3$ that we
assume for our filter, $\sigma_{\rm \sss G}=22\,(29)$ arcsec
respectively.

Figure \ref{fig:profiles} shows the comparison of the theoretical and
measured profiles for the unfiltered convergence maps of the fiducial
model corresponding to sources at redshift 1, in the absence of shape
noise. We use the hierarchical method to assign masses to peaks. The
peaks are binned according to the assigned mass, and the coordinates
of their centres are used to measure the shear and convergence
profiles. Each panel in the figure corresponds to a mass bin. The red
points depict the average of the measured convergence profiles of the
detected peaks. The error bars correspond to errors on the mean of the
128 fiducial fields. The solid blue lines represent the theoretical
profile of \eqn{eq:kmodel2}, estimated for the mean mass of the peaks
in the bin and the redshift $\zm=0.3$, i.e. the optimal redshift for
lensing for sources at redshift 1. The agreement between the model and
measurements is remarkable, given the fact that some peaks correspond
to halos at different redshifts or to no halos at all, and the fact
that we assume a spherical density model, which is bound to fail for
peaks arising from aspherical halos, or to be affected by projection
effects. Despite these limitations, the hierarchical method classifies
peaks efficiently on average, as shown in the figure. Note that the
bottom panels of Figure \ref{fig:profiles} present an oscillatory
feature around the virial radius of the profiles. We generated
convergence maps of synthetic, perfect NFW halos, and checked that
such features can appear if noise is added to the maps. This owes to
the fact that the compensated filter prefers to select peaks which
have regions of low convergence around the virial radius. This effect
is more pronounced for smaller peaks because these most likely
correspond to small halos, which have increased particle shot
noise. We have measured the convergence profiles of the
friends-of-friends (FoF) halos of the simulations at redshift 0.3, and
did not find such features as seen in Figure \ref{fig:profiles}, from
which we conclude that they are caused by the filter selection.

We use the above-mentioned values for $\alpha$ in \eqn{eq:kmodel2} to
obtain Figure \ref{fig:profiles}; based on several trials, we find
these values to yield the closest resemblance between the measured and
theoretical profiles. We do this test in the absence of shape noise,
since we are trying to address a technical issue arising from our
numerical simulations: the impact of the softening length. In the
absence of shape noise, the algorithm in \S\ref{III2} can be applied
by formally setting $\shn\rightarrow 1$ in \eqn{eq:filterC_def}, and
using only the mass criterion to select peaks; \eqn{eq:smoothed_map2}
does not change.

The hierarchical method could be useful for determining cluster masses
from WL profiles. In order to increase the accuracy of the results,
one should perform a careful analysis of: the chosen filter and its
parameters, the compensation radius $\theta_A$, the inclusion of the
halo-matter cross-correlation term visible in Figure
\ref{fig:profiles}, the impact of shape noise and projection noise,
the impact of photometric redshifts errors of the source galaxies.
However, this is beyond the goals of the present study.
\section{Results}
\label{IV}
\begin{figure*}
\centering
\includegraphics[scale=0.70]{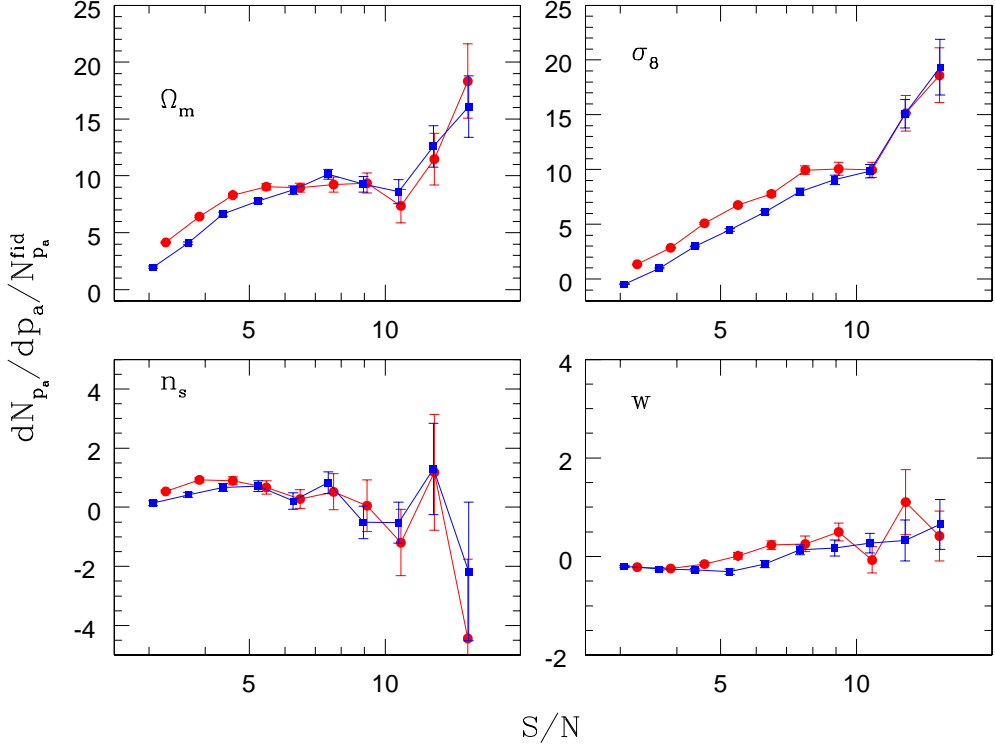}
\caption {Derivatives of the peak abundances with respect to the
  cosmological parameters considered, as a function of $\SN$. The red
  solid circles are measurements with the hierarchical method, and the
  blue solid squares correspond to smoothing with a filter of fixed
  size $M=2\times 10^{15} \Msol$ (angular size of the radius 13.2
  arcmin for $\zm=0.3$). The derivatives are estimated in accord with
  \eqn{eq:derivs}, and are divided by the mean counts of the fiducial
  model. We show the errors on the mean of 64 fields.}
\label{fig:derivs_sn}
\end{figure*}
We present a comparison between peak statistics results obtained
through the hierarchical algorithm described in \S\ref{III} and from
applying three single-sized filters of different size. When using the
single-sized filters, we keep the same filter function as given in
section \S\ref{III}, as this work is not concerned with assessing the
performance of filters of different shape. In this case, we simply
select the peaks by requiring that their $\SN \geq (\SN)_{\rm min}$,
with the $\SN$ given by \eqn{eq:SNc}. The three sizes that we consider
correspond at redshift $\zm=0.3$ to the masses $\{2\times 10^{15},
5.5\times 10^{14}, 10^{14}\}\, \Msol$, with the angular size of the
virial radii given by $\{13.2,\, 8.6,\, 4.8\}$ arcmin,
respectively. Note that due to the fact that ST halos have larger
radii than NFW ones of the same mass, these angular sizes are also
slightly larger than NFW angular sizes. For the hierarchical filter we
consider a series of 12 filters, logarithmically spanning the mass
interval $[8.85 \times 10^{13}, 2\times10^{15}]\,\Msol$, and with $\SN
\geq 2.6$. Ultimately, throughout the entire analysis for the
fixed-size and hierarchical methods, we shall use only peaks above the
threshold $(\SN)_{\rm min}=2.8$, but going to lower values ensures the
completeness of the sample of hierarchical peaks. The shape noise
contamination makes it difficult to consider smaller filters. We bin
the resulting peak abundances in terms of $\SN$, logarithmically
spanning an interval $[2.8, 14]$. For reasons discussed in Appendix
\S\ref{II}, we choose $N_{\rm bin} = 20$. Note that for the
hierarchical abundance it is useful to also consider binning in mass,
as assigned through
Eqs~(\ref{eq:smoothed_map2}),~(\ref{eq:match}),~(\ref{eq:Mtrue}). This
allows to draw analogies between the properties of the WL peaks and
those of 3D haloes. Here too we use 20 bins spanning $[10^{14}, 2
  \times 10^{15}]\,\Msol$, the lower bound roughly corresponding to
the $(\SN)_{\rm min}=2.8$ for the analyzed cosmological models.

Comparing peak abundances obtained with different methods is not
necessarily relevant: the results will be clearly different, and it
would be hard to decide which filter size is more effective. This is
shown in Figure \ref{fig:pfSN}, where we present the peak functions
corresponding to the three single-sized filters, as well as the
hierarchical method. The functions are expressed as number of peaks
per unit $\SN$ for an area of $144\,\deg2$, and the results are an
average of the peak functions measured in the 128 fields of the
fiducial model. The error bars correspond to errors on the mean. As
expected, each single-sized filter favors the detection of peaks with
$\SN$ in accord to its size: the smallest filter peak abundance is
mostly formed by low $\SN$ peaks, and similarly for the medium and
large filters. The hierarchical peak abundance is similar to the
largest-filter abundance for the high-$\SN$ bins, and to the
smallest-filter abundance for the low-$\SN$ bins. We shall next
explore how the measured abundances translate into cosmological
constraints.

To this effect, we shall resort to the Fisher matrix formalism, with
four clear goals: \begin{enumerate} \item to provide a comparison
  between the filtering methods; \item to test which range of mass or
  $\SN$ contributes most to the constraints derived from WL peak
  counts; \item to test the difference between the errors obtained by
  using the full covariance matrix of counts, and the Poisson
  errors; \item to provide a realistic forecast for surveys like
  $\lsst$ and $\euclid$, in a very direct manner, based on simulation
  measurements. \end{enumerate}
\subsection{Fisher matrix considerations}
\label{IV1}
\begin{figure*}
\centering
\includegraphics[scale=0.7]{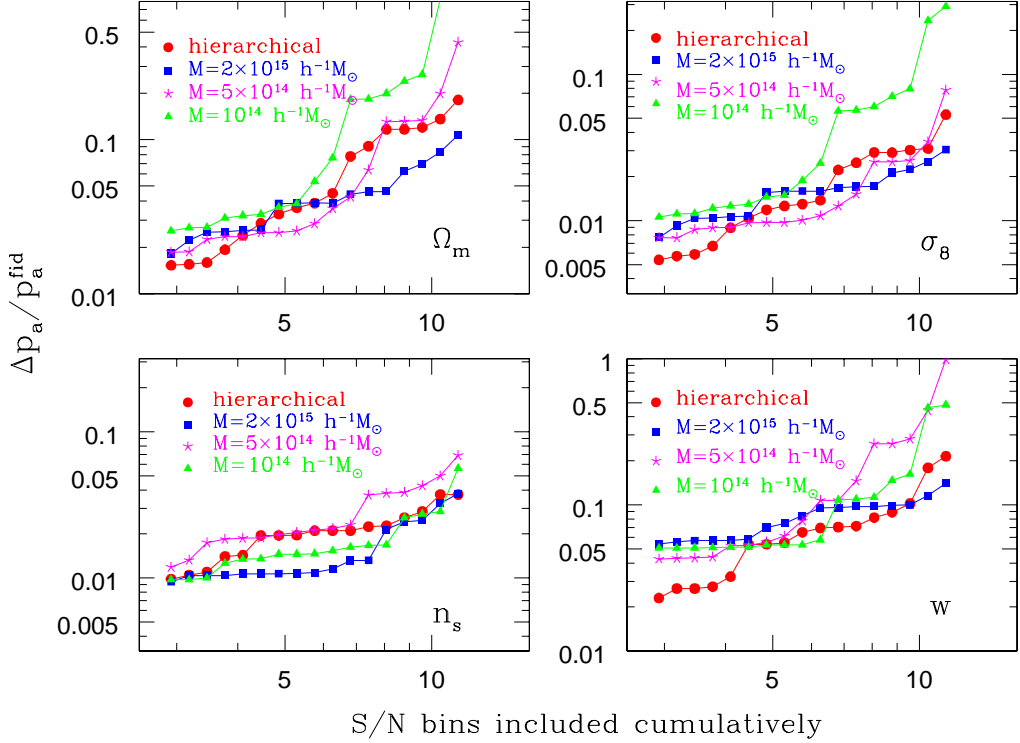}
\caption {Fractional marginalized Fisher matrix errors based on
  measurements from simulations. The symbols and colors are the same
  as in Figure \ref{fig:pfSN}. The errors are cumulative: starting
  from the highest bins, we gradually allow the rest of the bins to
  contribute to the constraints. The central values of the $\SN$ bins
  are indicated on the ${\bf\hat x}$-axis. }
\label{fig:cum_errorsM_sn}
\end{figure*}
Using the measured peak abundances, we compute the Fisher information
following the standard definition:
\be 
\Fish_{p_ap_b} = - \left< \frac{\partial^2 \ln {\cal L}}{\partial p_a\partial p_b}\right>,
\label{eq:Fisher_gen}
\ee
where $p_{a}$ and $p_b$ are elements of the cosmological model
parameter set $\mx{\bm$p$}$ upon which the likelihood ${\mathcal L}$
depends. In our case the set is: $\{n,\,\s8,\,\om,\, w\}$.
We assume a Gaussian likelihood
\ba
{\cal L}(\m|\,\bar{\m}(\p), \C(\p))=\frac{1}{(2\pi)^{N_{\rm bin}/2}|\C|^{1/2}}
\hspace{2.6cm} \nn \\ \times \exp\left[-\frac{1}{2}(\m-\bar{\m})^{t}\C^{-1}
(\m-\bar{\m})\right],
\ea
where $\m$ is the vector of peak counts, and $\bar{\m}$ is the vector
of mean number of peaks; both vectors have the dimension $N_{\rm
  bin}$, i.e. the number of bins considered. The covariance matrix of
the counts in bins $i$ and $j$ is
\be
{\rm C}_{i\,j}=\langle (m_i-\bar m_i)\,(m_j-\bar m_j)\rangle .
\label{eq:cov}
\ee
From the Fisher matrix, one may obtain an estimate of the marginalized
errors and covariances of the parameters:
\be 
\sigma^2_{p_ap_b} = [\Fish^{-1}]_{p_ap_b},
\label{eq:marg_error}
\ee
as well as the unmarginalized errors: 
\be \sigma_{p_a} = [\Fish_{p_ap_a}]^{-1/2} .
\label{eq:unmarg_error}
\ee
%
The size of the errors quantifies the efficiency of the filtering
method to extract cosmological information from WL peak counts. For
simplicity, we shall ignore the trace term in the Fisher matrix,
\citep{Tegmarketal1997}. In this case, \eqn{eq:Fisher_gen} can be
rewritten as
\be
\Fish_{p_a\,p_b} = \sum_{i, j} \frac{\partial {\bar m_i}}{\partial p_a}\, 
{\rm C}^{-1}_{i\,j}\,\frac{\partial {\bar m_j}}{\partial p_b},
\label{eq:fisher_counts}
\ee
We are also interested in the Poisson errors of the peak counts, since
the Poisson statistic is widely adopted in forecasting cosmological
constraints from WL peak counts. They are given by
\be
\Fish^P_{p_a\,p_b} = \sum_{i} \frac{\partial {\bar m_i}}{\partial p_a}\, 
\frac{\partial {\bar m_i}}{\partial p_b}\,\frac{1}{\bar m_i}.
\label{eq:fisher_poisson}
\ee
The mean number of counts for bin $i$ is estimated as
\be
\hat{\bar m}_i=\frac{1}{N}\sum_{f=1}^{N} m_i^f
\ee
In the above $f$ designates the field number, while $N$ is the total
number of fields; for the fiducial cosmology, $N=128$, and for the
variational cosmologies $N=64$. An unbiased, maximum-likelihood
estimator for the covariance matrix is:
\be
\hat{\rm C}_{i j}=\frac{1}{N-1}\sum_{f=1}^{N} (m_i^f-\hat{\bar m}_i)
\,(m_j^f-\hat{\bar m}_j)
\label{eq:cov2}
\ee
The derivatives of the counts with respect to the cosmological
parameters are calculated from
\be
\widehat{\frac{\partial \bar m_i}{\partial p_a}}=\frac{1}{N}\sum_{f=1}^{N} 
\frac{m_i^f(p_a+\Delta p_a)-m_i^f(p_a-\Delta p_a)}{2\Delta p_a},
\label{eq:derivs}
\ee
where $\Delta p_a$ represents the $\pm$ step in the cosmological
parameters, e.g. Table \ref{tab:zHORIZONcospar}. 

We estimate the Fisher matrix errors using the covariance on the mean
for the counts of the fiducial model; the rescaled covariance matrix
corresponds to an area of $\approx 18000\,\deg2$. Together with the
survey specifications given in section \S\ref{II}, this makes our
study representative for two future surveys, $\lsst$ and $\euclid$.
\subsection{Comparison of filtering methods}
\label{IV4}
Figure \ref{fig:derivs_sn} depicts the derivatives of the measured
peak abundances with respect to the four cosmological parameters that
we consider, as a function of $\SN$. We show results for the
hierarchical method and the largest of the single-sized filters,
$M=2 \times 10^{15} \Msol$. The derivatives are estimated using
\eqn{eq:derivs}, and the result is divided by the mean counts of the
fiducial cosmology. The figure shows that both filtering methods yield
peak functions similarly sensitive to cosmology, with the hierarchical
derivatives displaying slightly more features than the single-sized
ones. For most of the $\SN$ range considered, the peak-function
derivatives are non-zero, signifying that there is cosmological
information in the high-$\SN$ peaks, as well as in the low-$\SN$ ones,
as previously noticed by \cite{DietrichHartlap2010}. This originates
in a similar behaviour displayed by the halo mass function: In a
previous work \citep{SmithMarian2011}, we found the derivatives of the
latter with respect to the same parameters studied here to be non-zero
for a large range of halo masses, down to $M=10^{13} \Msol$
(compare Figures 6, 7 in that work with Figures \ref{fig:derivs_sn},
\ref{fig:cum_errorsM_sn} in this work).

Figure \ref{fig:cum_errorsM_sn} depicts the marginalized Fisher errors
for the four cosmological parameters that we consider. The errors are
fractional, i.e. the error for each parameter is divided by the
fiducial value of that parameter, and cumulative, i.e. we include
$\SN$ bins cumulatively, from the largest to the lowest. The central
values of the bins are indicated on the ${\bf\hat x}$-axis. The
symbols and colors are the same as in Figure \ref{fig:pfSN}. It is
apparent that the hierarchical filtering performs better than the
single-sized filtering, if one takes into account peaks with $\SN\leq
4$. The greatest improvement is for $w$: the hierarchical error is
smaller by more than a factor of 2 compared to the fixed-size
filtering. A smaller improvement happens also in the case of $\s8$ and
$\om$, while the error on $\ns$ seems unaffected by the filtering
method, due to generally poor constraining power that peaks have on
this parameter. Figure \ref{fig:cum_errorsM_sn} reinforces the
suggestion of Figure \ref{fig:derivs_sn} that the inclusion of peaks
with small $\SN$ improves significantly the cosmological information.

Note that for the hierarchical method we can also use aperture-mass
bins to measure the peak function derivatives, the covariance matrix,
and the Fisher matrix, with the mass given by \eqn{eq:Mtrue}. We
obtain similar results to those presented in Figures
\ref{fig:derivs_sn}, \ref{fig:cum_errorsM_sn}. The single-sized
filters perform very similarly, the largest one being marginally
better in the case of $\om$, $\s8$, and $\ns$. Its diameter of 13.2
arcmin is larger than what previous studies in the literature have
used: \cite{Hamanaetal2004} had a $1$-arcmin Gaussian filter,
\cite{HennawiSpergel2005} employed a $\sim 5$-arcmin NFW filter, and
\cite{DietrichHartlap2010} a $5.6$-arcmin one. Since the constraints
from filtering with fixed sizes are so similar, we shall only show the
results from the marginally-better $M=2\times 10^{15} \Msol$ one.

In Appendix \S\ref{AI} we examine the statistical properties of the
peaks detected through the hierarchical and fixed-sized methods. We
find that: \begin{itemize} \item The Poisson statistic describes well
  the distribution of hierarchical high-$\SN$ and high-mass peaks. The
  mass is defined by \eqn{eq:Mtrue}. We show that the correlation
  matrix of $\SN$-binned hierarchical peaks has strong off-diagonal
  contributions for the small-$\SN$ bins, while being largely diagonal
  for the large-$\SN$ bins. The same applies to the mass-binned
  correlation coefficient, and this behaviour is similar to that of
  halos, as shown in \cite{SmithMarian2011}.
\item The high-$\SN$ single-sized peaks are also reasonably described
  by the Poisson distribution, due to the fact that such peaks are
  usually quite massive and rare. The correlation matrix of these
  peaks seems slightly more correlated for $\SN \geq 7 $ than the
  hierarchical matrix.
\end{itemize}
\begin{figure*}
\centering
\includegraphics[scale=0.65]{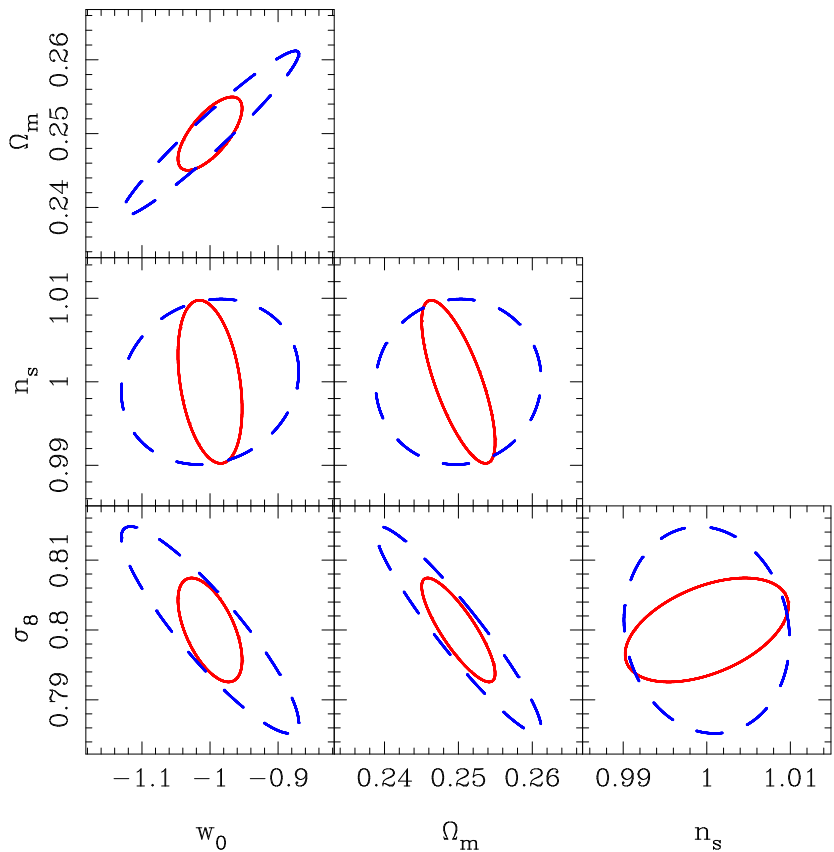}
\caption {Forecasted marginalized errors for WL peak counts from a
  \euclid-type of survey combined with CMB constraints from
  \planck. The blue dashed ellipses depict the results of the fixed
  filter $M=2 \times 10^{15}\, \Msol $, while the red solid ellipses
  correspond to the hierarchical method. In both cases, the WL Fisher
  matrix has been added to the Planck Fisher matrix, and the
  constraints are at the $2-\sigma$ level.}
\label{fig:ellipses}
\end{figure*}
Lastly, in Appendix \S\ref{AII} we investigate the dependence of the
Fisher errors on the number of bins in which the $\SN$ interval is
divided. Figure \ref{fig:SNbin_dependence} suggests that all filtering
methods reach the expected saturation in information if $N_{\rm bin}
\leq 20$, which is why we choose $N_{\rm bin}=20$ for the results
presented in this work.
\subsection{ Forecasting constraints on cosmology}
\label{IV5}
\begin{table*}
\caption{Fisher matrix constraints for the hierarchical method and a
  fixed filter of $M=2 \times 10^{15}\,\Msol$. The fiducial values for the
  parameters are ${\ns=1,\,\om=0.25,\,\s8=0.8,\,w=-1}$.
\label{tab:Fisher_constraints}}
\vspace{0.2cm}
\centering{
\begin{tabular} {c}
\hspace{3.5cm} Hierarchical errors \hspace{2.3cm} Fixed-size errors\\
\begin{tabular}{c|cccccccc}
\hline 
 &  $\ns$ & $\om$ & $\s8$ & $w$ & $\ns$ & $\om$ & $\s8$ & $w$  \\
{\tt Unmarginalized}  & 0.0025 \ &  0.0006 & 0.0015 & 0.0125 & 0.0064 & 0.0008 & 0.0019 & 0.0128 \\
{\tt Marginalized}  & 0.0094 \ &  0.0038 & 0.0043 &  0.0235 & 0.0105 & 0.0046 & 0.0061 & 0.0552 \\
{\tt Marginalized + CMB}  & 0.0039 \ &  0.002  & 0.003 &  0.019 & 0.004 & 0.0045 & 0.006 & 0.0525\\
\hline
\end{tabular}
\end{tabular}}
\end{table*}
We present a Fisher-matrix forecast for the 4-dimensional cosmological
parameter space explored in this work. This will enable us to compare
the filtering methods in a more realistic context, using marginalized
errors and also cosmic microwave background (CMB) information. 

For the Planck Fisher matrix, we shall assume that the CMB temperature
and polarization spectra can constrain 9 parameters: the dark energy
equation-of-state parameters $w_0$ and $w_a$; the density parameter
for dark energy $\Omega_{\rm DE}$; the CDM and baryon density
parameters scaled by the square of the dimensionless Hubble parameter
$\omega_{\rm CDM}=\Omega_{\rm CDM}h^2$ and $\omega_{\rm b}=\Omega_{\rm b}h^2$
($h=H_0/[100\, {\rm km\,s^{-1}\,Mpc^{-1}}]$); the primordial spectral
index of scalar perturbations $\ns$; the primordial amplitude of
scalar perturbations $A_s$; the running of the spectral index
$\alpha$; and the optical depth to the last scattering surface
$\tau$. To compute the CMB Fisher matrix we follow
\cite{Eisensteinetal1999}:

\be 
\Fish_{p_a\,p_b}=\sum_l \sum_{X,Y} \frac{\partial C_{l,X}}{\partial p_a}
{\rm Cov}^{-1}\left[C_{l,X},C_{l,Y}\right]
\frac{\partial C_{l,Y}}{\partial p_b}\ ,
\ee
where $\{X,Y\}\in \{{\rm TT},\,{\rm EE},\,{\rm TE},\,{\rm BB}\}$,
where $C_{l,\rm TT}$ is the temperature power spectrum, $C_{l,\rm EE}$
is the E-mode polarization power spectrum, $C_{l,\rm TE}$ is the
temperature-E-mode polarization cross-power spectrum, and $C_{l,\rm
  BB}$ is the B-mode polarization power spectrum. The assumed sky
coverage is $f_{\rm sky}=0.8$ In order to make the CMB Fisher matrix
compatible with our parameters, we rotate it to a new set
\be 
{\bf q}^{T}=\{w_0, w_a, \Omega_{m}, h, f_b, \tau, \ns, \s8, \alpha \} \ ,
\ee
where for us $w_0=w$. We marginalize over the 5 parameters absent
from our analysis.

Table \ref{tab:Fisher_constraints} and Figure \ref{fig:ellipses}
represent the main results of this work, showing the overall
improvement the hierarchical method brings over the fixed-size method
after marginalization and especially after the inclusion of the CMB
information. For the fixed-size method we choose the filter with
$M=2\times 10^{15}\,\Msol$, i.e. the best-performing filter among the
fixed sizes that we have probed. We also show the unmarginalized
errors, for a more complete picture.

Combined with the CMB, the hierarchical errors are a factor of 2
better than the single-filter method for $\om$ and $\s8$, and almost a
factor of 3 better for $w$. For $\ns$ there is no significant
difference between the filtering methods. This happens because the CMB
constrains the primordial power spectrum tighter than WL peak counts.

We further depict these results in Figure \ref{fig:ellipses} as
$2\sigma$-ellipses; the blue dashed ellipses correspond to the
single-sized method, and the red solid ones to the hierarchical
algorithm. Here we see again that the latter really improves the joint
constraints for $\{\s8, \om, w\}$.

It is difficult to make a comparison to previous forecasts in the
literature, as the probed parameter space and survey specifications
are not the same, so we shall mention only two. \cite{Wangetal2004}
presented a forecast for $\lsst$ in which besides WL counts they also
include the cluster power spectrum, which they treat as completely
independent. They consider a larger parameter space than ours,
including $w_a$ and $\omega_b$, and assume $\nbar =65\,\arcmint^{-2}$
and $(\SN)_{\rm min}=4.5$. The constraints that they find when
combined with \planck \hspace{0.07cm} priors (see Table 6 in their
paper) are: $\Delta \ns=0.0022,\,\Delta \Omega_{\rm DE}=0.0033,\,
\Delta \s8=0.0037, \Delta w=0.036$. They use a Gaussian filter of
size $1$ arcmin, and their fiducial model has $\om=0.27, \,
\s8=0.9$. The results are rather similar to ours, though their
constraint of $w$ is surprisingly tight, given the sensitivity of
$\Delta w$ to $(\SN)_{\rm min}$ -- higher than ours -- and the fact that
they include $w_a$, known to degrade substantially the constraint on
$w$.

We also make a comparison to our previous work
\citep{MarianBernstein2006}, which uses the same type of normalized
filter as this study. The detection threshold is 5, the projection
noise is accounted for, and instead of $\ns$, $w_a$ is considered. The
fiducial values for $\om$ and $\s8$ are the same as in
\cite{Wangetal2004}; combining with the Planck information, we found
the constraints: $\Delta \om=0.005,\, \Delta \s8=0.004, \Delta
w=0.063$. These are in agreement with the results from the present
study, given the above-mentioned differences.
\section{ Summary and conclusions }
\label{V}
In this paper we proposed a new method, which we called `the
hierarchical algorithm' to detect and explore WL peak counts. While
previous studies have examined the benefits of using filters of a
certain shape \citep{HennawiSpergel2005, Maturietal2005,
  Gruenetal2011}, here we have focused on the way the filtering should
be performed to maximize the inferred cosmological constraints. To
this goal, we have used a large set of WL maps produced by ray-tracing
through $N$-body simulations with varying cosmological models, as
described in section \S\ref{II}.

Our method was based on the idea of sequential smoothing of the maps
with filters of different size, from the largest to the smallest. The
chosen filter was an aperture-mass filter, matching the NFW density
profile of halos. Combining the information contained in the maps
smoothed on different scales, we determined the largest filter size
for which a peak would not only be a point of local maximum, with a
$\SN$ larger than a certain threshold, but also a match to the NFW
profile of the filter. For the latter we have assumed a fixed redshift
equal to the optimal redshift for lensing given the mean redshift of
the source distribution. Under this assumption, we assigned a unique
value of mass and $\SN$ to the detected peaks, as described in detail
in section \S\ref{III}. Thus, the peak function arising from the
hierarchical method does not depend on a particular filter size.

We compared the hierarchical peak abundance to that obtained from
applying a filter of fixed size; for the latter we used the same
aperture filter and considered three sizes $\{13.2,\,8.6,\,4.8\}$
arcmin. At the assumed redshift $\zm=0.3$, these correspond
to halos with a Sheth-Tormen mass of $M=\{2\times10^{15},
5\times10^{14}, 10^{14}\}\, \Msol$. To quantify the efficiency of the
smoothing methods, we took the Fisher matrix approach: we compared the
errors on the cosmological parameters derived from each method. The
considered parameters were: $\{\ns, \om, \s8, w\}$. Our
findings are as follows:
\begin{enumerate} 
\item The marginalized Fisher matrix errors obtained from the
  hierarchical peak abundance combined with CMB information from
  $\planck$ were better by a factor of $\approx 2$ compared to the
  results of the single-sized filtering. This was true if we took into
  account low peaks with $\SN \sim 3$; if we allowed only peaks with
  $\SN\geq 6$ then the hierarchical errors were only marginally
  better.
\item The three filters of fixed size yield very similar results, the
  largest being slightly more effective.
\item We have provided a cosmology forecast for WL peak counts
  relevant to future surveys like $\euclid$ and $\lsst$. Combined with
  information from a CMB experiment such as $\planck$, the
  hierarchical marginalized errors for the considered parameters were:
  {$\Delta \ns=0.0039\, (0.004);\, \Delta \om=0.002\, (0.0045);\,
    \Delta \s8=0.003\, (0.006);\, \Delta w=0.019\, (0.0525)$}, where
  the values in the parenthesis corresponded to the results of the
  largest, fixed-size filter. Note that we have assumed no knowledge
  of the redshifts of the peaks, and yet have obtained values in a
  reasonable accord with analytical forecasts in the literature.
\item The high-$\SN$ and high-mass ends of the hierarchical peak
  function were reasonably described by the Poisson distribution,
  e.g. Figures~\ref{fig:corrcoefHm},~\ref{fig:corrcoefHsn},
  ~\ref{fig:poisson_sn_H_B},~\ref{fig:poisson_m_H}, since the
  hierarchical filtering successfully assigned the largest mass and
  $\SN$ to the largest and rarest peaks.
\item The results of the Fisher matrix analysis had a slight
  dependence on the number of $\SN$ bins used: The most suitable
  number of bins for the hierarchical method was 20. 
\end{enumerate}
We have checked that the hierarchical method yields similar
constraints if one bins the peak information in mass and not $\SN$,
which is a reassuring consistency check.

There are certain improvements that one could bring to the
hierarchical method. First, the choice of filter shape: in this study,
we have resorted to a filter which is optimal if one assumes the shape
noise of galaxies as the main source of noise for WL
measurements. Though this filter is also effective in reducing the
impact of correlated line-of-sight projections for the measured peaks
\citep{MarianSB2009, MarianSB2010}, one could use a more sophisticated
shape, as discussed in \cite{Gruenetal2011}. Second, one should test
the benefits of having more redshift information on the source
galaxies, i.e. use tomography to improve the cosmological constraints
derived from the peak abundance. We defer these issues to a future
study.

The main message conveyed by our work is that, compared to the
standard approach of single-sized smoothing usually discussed in the
literature, the hierarchical method extracts significantly more of the
cosmological information enclosed in WL peak counts. Therefore, it
will be a very useful tool for surveys like $\euclid$ and $\lsst$
which have the potential to detect many thousands of peaks.

\subsection*{Acknowledgements}
We thank Gary Bernstein for his comments on the manuscript. We also
thank V. Springel for making public {\tt Gadget-2} and for providing
his B-FoF halo finder. LM, SH, and PS are supported by the Deutsche
Forschungsgemeinschaft (DFG) through the grant MA 4967/1-1, through
the Priority Programme 1177 `Galaxy Evolution' (SCHN 342/6 and WH
6/3), and through the Transregio TR33 `The Dark Universe'. SH also
acknowledges support by NSF grant number AST-0807458-002. RES was
partly supported by the Swiss National Foundation under contract
200021-116696/1, the WCU grant R32-2008-000-10130-0, and the
University of Z\"{u}rich under contract FK UZH 57184001. RES also
acknowledges support from a Marie Curie Reintegration Grant and the
Alexander von Humboldt Foundation.
\bibliographystyle{mn2e}

\appendix
\label{A}
\section{The correlation coefficient of the peak counts}
\label{AI}

\begin{figure}
\centering
\includegraphics[scale=0.45]{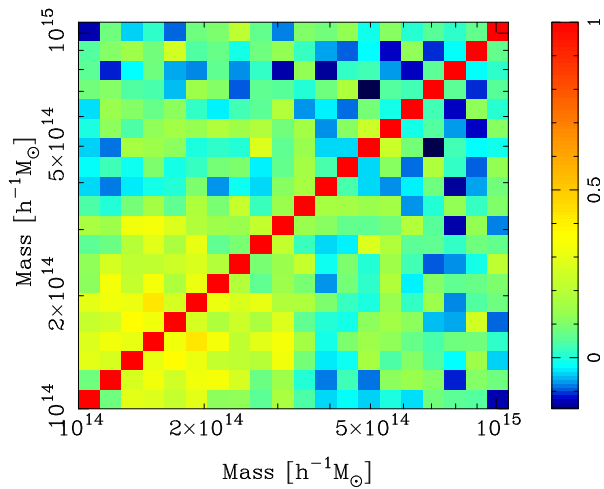}
\caption {The correlation coefficient $\rr_{ij}$ for the hierarchical
  peak abundance, binned in mass. The measurements are an average of
  128 fields of the fiducial model.}
\label{fig:corrcoefHm}
\end{figure}
\begin{figure}
\centering
\includegraphics[scale=0.45]{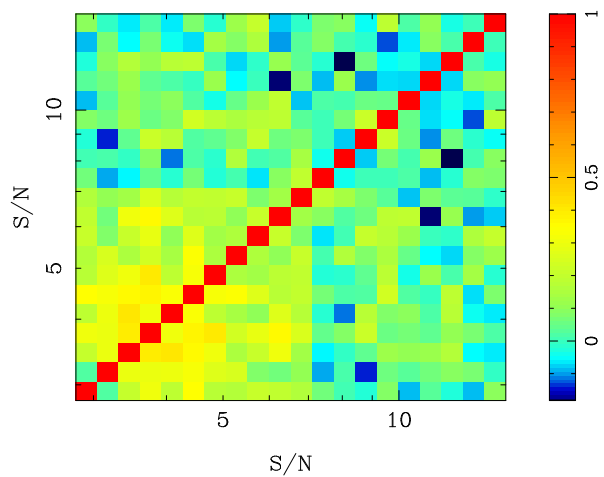}
\caption {Same as in the previous figure, only the binning is in
  $\SN$.}
\label{fig:corrcoefHsn}
\end{figure}
\begin{figure}
\includegraphics[scale=0.45]{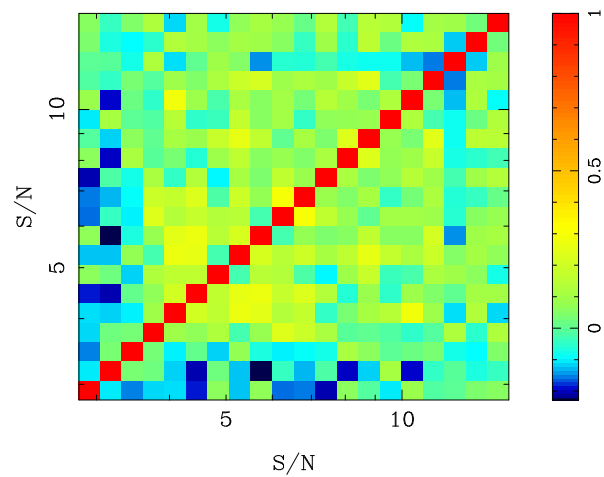}
\caption {The correlation coefficient for the fiducial peak abundance
  measured with a filter of fixed size $M=2 \times 10^{15} \Msol$. The
  binning is in $\SN$.}
\label{fig:corrcoefBsn}
\end{figure}
\begin{figure}
\centering
\includegraphics[scale=0.44]{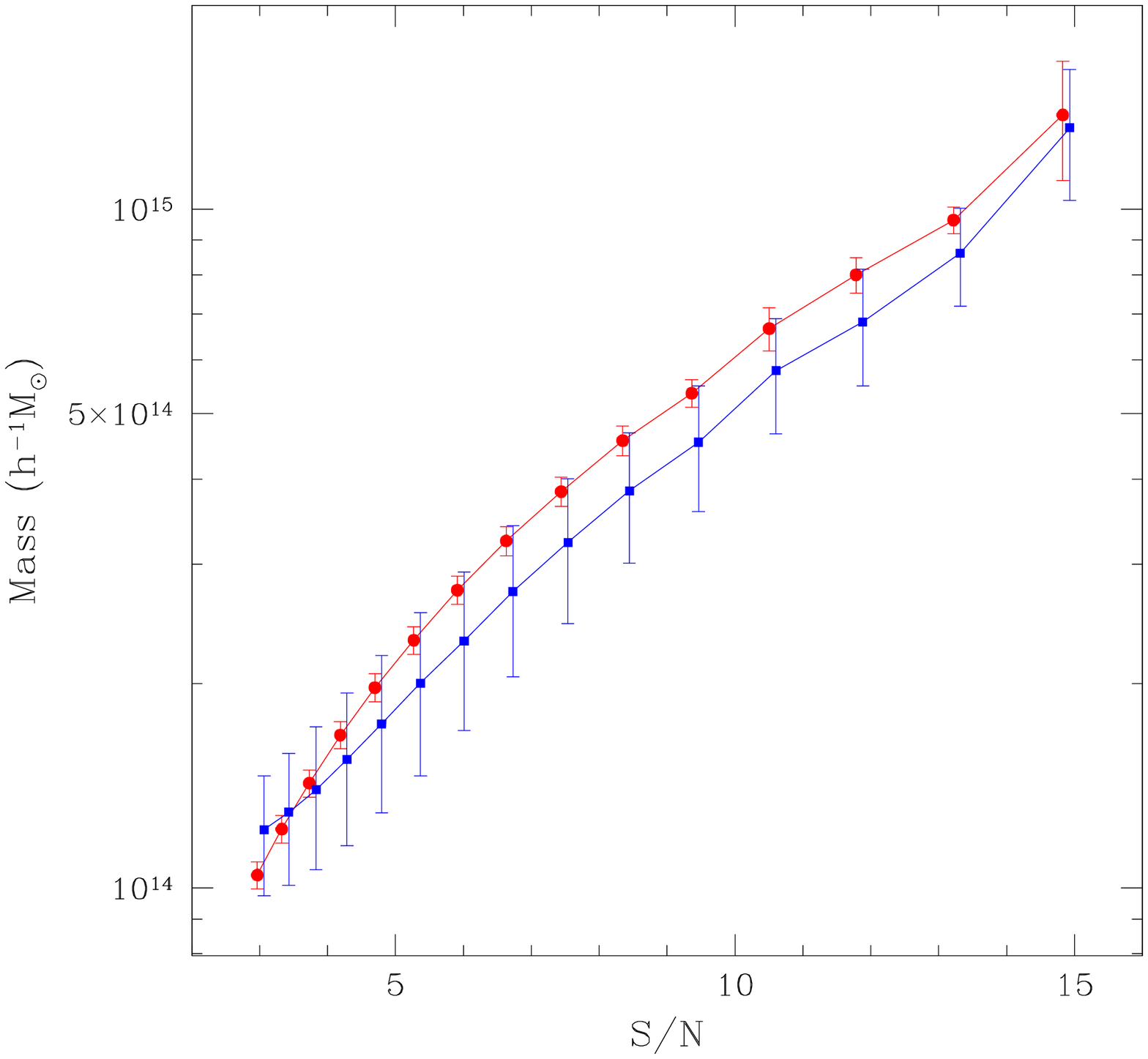}
\caption{The $\SN$-mass relation for the hierarchical (red solid
  circles) and single-sized filtering methods (blue solid
  squares). For the latter, we adopted the size $M=2 \times 10^{15}\,
  \Msol$. Only peaks detected through both methods are being
  compared. For both methods, the mass is {\it the hierarchical
    aperture mass} of \eqn{eq:Mtrue}. For the hierarchical method the
  relationship between the assigned mass and $\SN$ of the peaks is
  very tight, with little scatter. For the single-sized method there
  is significantly more scatter, particularly for the lower $\SN$
  bins. For clarity, we have slightly offset the position of the
  points on the {\bf x}-axis.}
\label{fig:scatter}
\end{figure}
For a covariance matrix $\C$, the correlation coefficient is defined by
\be
{\rm r}_{ij}=\frac{{\rm C}_{ij}}{\sqrt{{\rm C}_{ii}\,{\rm C}_{jj}}}, 
\ee
We explore the correlation coefficient of the peak abundance based on
measurements from 128 fields of the fiducial model. Figure
\ref{fig:corrcoefHm} shows $\rr$ for the hierarchical method, using
aperture mass bins. The mass is assigned according to the description
given in section \S\ref{III}. The result is very similar to the
correlation coefficient of the 3D halo abundance
\citep{SmithMarian2011}: There are strong correlations at the low-mass
end ($M<3\times10^{14} \Msol$) while the large-mass bins are mostly
uncorrelated. In Figure \ref{fig:corrcoefHsn} we show the hierarchical
results for $\rr$ based on the same measurements, but using bins in
$\SN$ instead of mass. The same pattern as in the previous figure is
visible, with the lower bins $\SN \leq 6$ having a correlation
coefficient of $\sim 0.4$. Finally, Figure \ref{fig:corrcoefBsn}
depicts $\rr$ measured from the same fiducial maps, using a single
filter of size $M=2 \times 10^{15} \Msol$. The correlations seem
weaker than for the hierarchical case, but they extend to higher
$\SN$, the highest bin being however largely uncorrelated. This is
further explained in the next figure.

Figure \ref{fig:scatter} presents the $\SN$-mass scatter of peaks
detected through the two methods. For the fiducial-cosmology peak
abundances, we compare the coordinates of the peaks, to select only
those peaks found through both methods. For these peaks, we
concentrate on the following 3 quantities: the hierarchical $\SN$
given by \eqn{eq:SNtrue}, the single-sized $\SN$ from \eqn{eq:SNs},
and the hierarchical aperture mass from \eqn{eq:Mtrue}. We consider
$\SN$ bins. For each bin we compute: the mean $\SN$ of the
hierarchical and single-sized peaks in that respective bin, as well as
the mean mass, and the error on the mass. We emphasize that for both
methods, by {\it mass} we mean the aperture mass assigned through the
hierarchical algorithm, i.e. \eqn{eq:Mtrue}. The figure shows that
both methods yield on average a similar relation between mass and
$\SN$. However, in the hierarchical case, this relation is very tight:
large-/small-mass peaks have large/small $\SN$, hence the similarity
between the correlation matrices in Figures~\ref{fig:corrcoefHm}
and~\ref{fig:corrcoefHsn}. For the single-filter method this is not
the case, as the size of the error bars suggests that peaks with quite
varying mass are binned in the same $\SN$ bin. This is consistent with
the correlation matrix shown in Figure~\ref{fig:corrcoefBsn}.

\begin{figure*}
\centering
\includegraphics[scale=0.62]{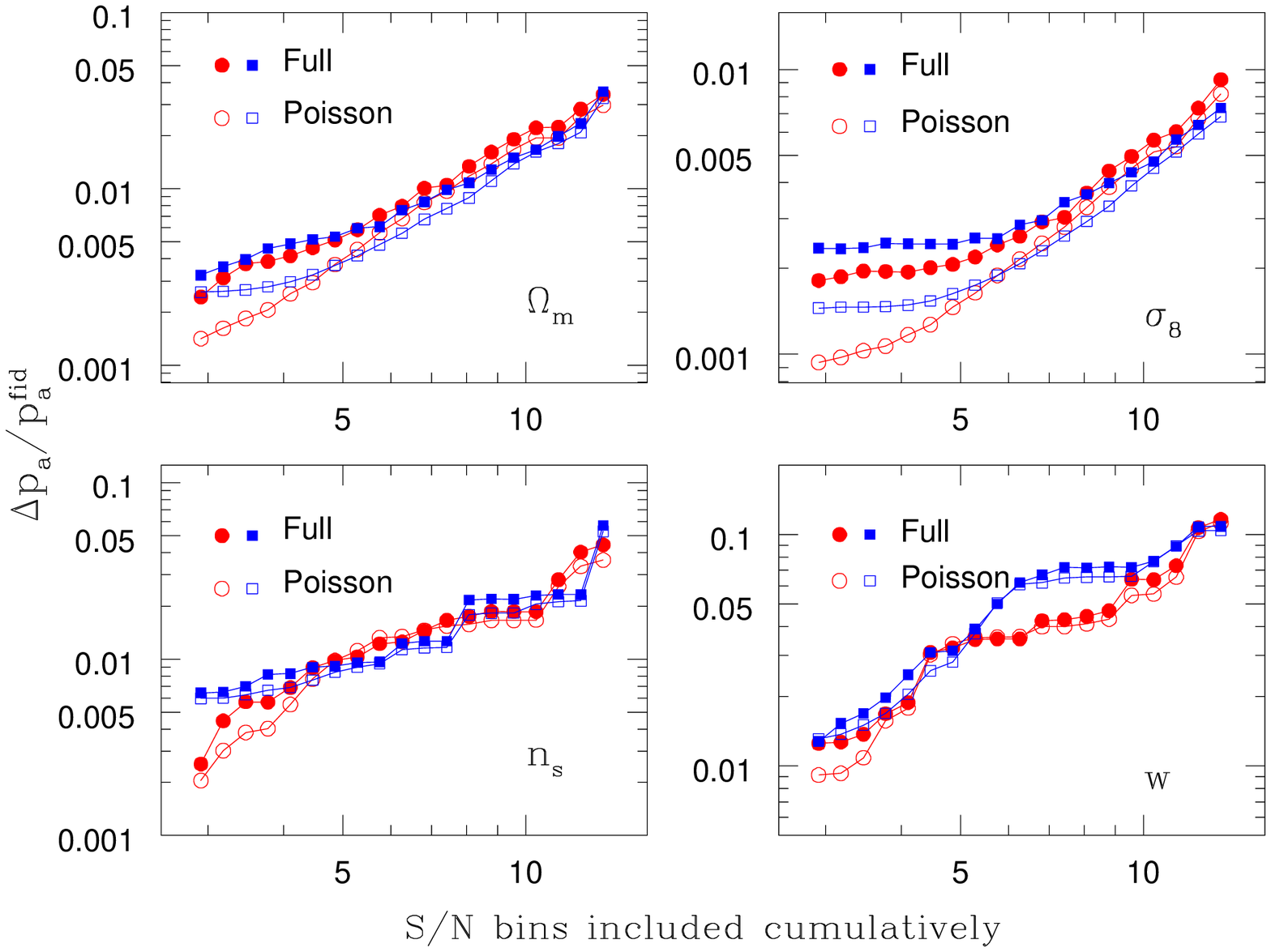}
\caption {Comparison between Poisson and full-covariance fractional
  unmarginalized errors,
  i.e. \eqns{eq:fisher_poisson}{eq:fisher_counts}. For clarity, we
  show only the results for a single-sized filter of $M=2 \times
  10^{15}\,\Msol$ -- blue squares -- and the hierarchical method --
  red circles. The full errors are the same as in Figure
  \ref{fig:cum_errorsM_sn}. }
\label{fig:poisson_sn_H_B}
\end{figure*}
\begin{figure*}
\centering
\includegraphics[scale=0.62]{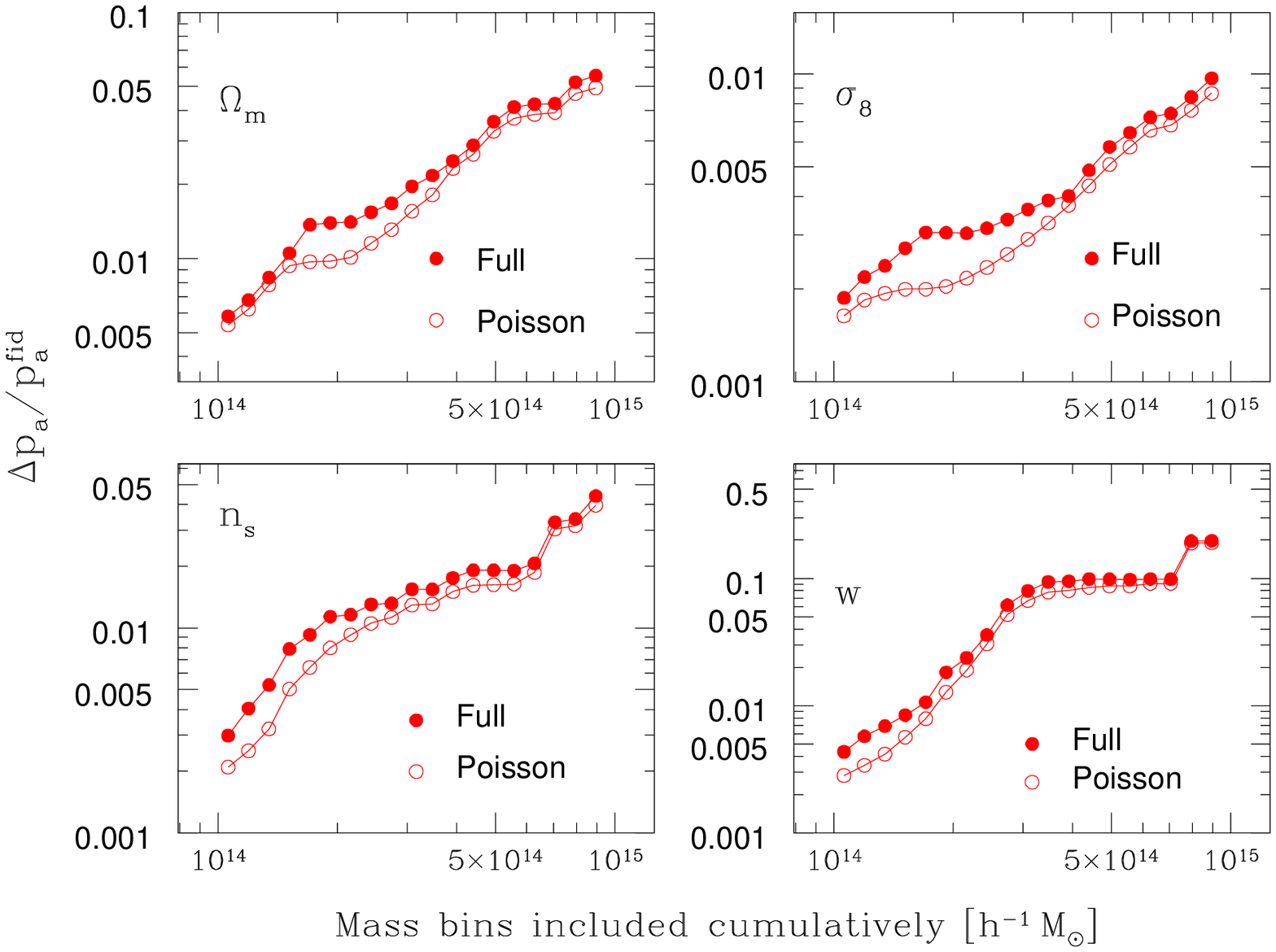}
\caption {Comparison between the full-covariance and Poisson
  cumulative unmarginalized errors, with the information binned in
  terms of aperture mass, i.e. \eqn{eq:Mtrue}. We show only results
  for the hierarchical method. The two errors converge at the
  high-mass end, so the hierarchical algorithm successfully assigns
  the highest masses to the most massive peaks, well described by the
  Poisson distribution. These results are similar to those shown in
  Figure \ref{fig:poisson_sn_H_B}.}
\label{fig:poisson_m_H}
\end{figure*}
In Figure \ref{fig:poisson_sn_H_B} we compare the fractional
cumulative errors obtained using the full covariance matrix
\eqn{eq:fisher_counts} to the Poisson errors \eqn{eq:fisher_poisson},
for the two methods considered. The Poisson and full-covariance errors
converge at the high-$\SN$ end, in accord with the fact that the most
massive peaks are assigned the largest $\SN$ in both the hierarchical
and single-filter methods. Finally, in Figure \ref{fig:poisson_m_H} we
show the fractional and cumulative Fisher errors for the hierarchical
method, considering binning in mass, and not $\SN$. As already
suggested by Figure \ref{fig:corrcoefHm}, the Poisson statistic
captures reasonably well the high-mass end of the distribution of
hierarchical peaks, where the full-covariance and Poisson errors
converge.

To conclude, the Poisson distribution {\it can} be used to approximate
the likelihood function of high-$\SN$ and high-mass hierarchical
peaks.
\section{ The number of bins}
\label{AII}
\begin{figure*}
\centering
\includegraphics[scale=0.7]{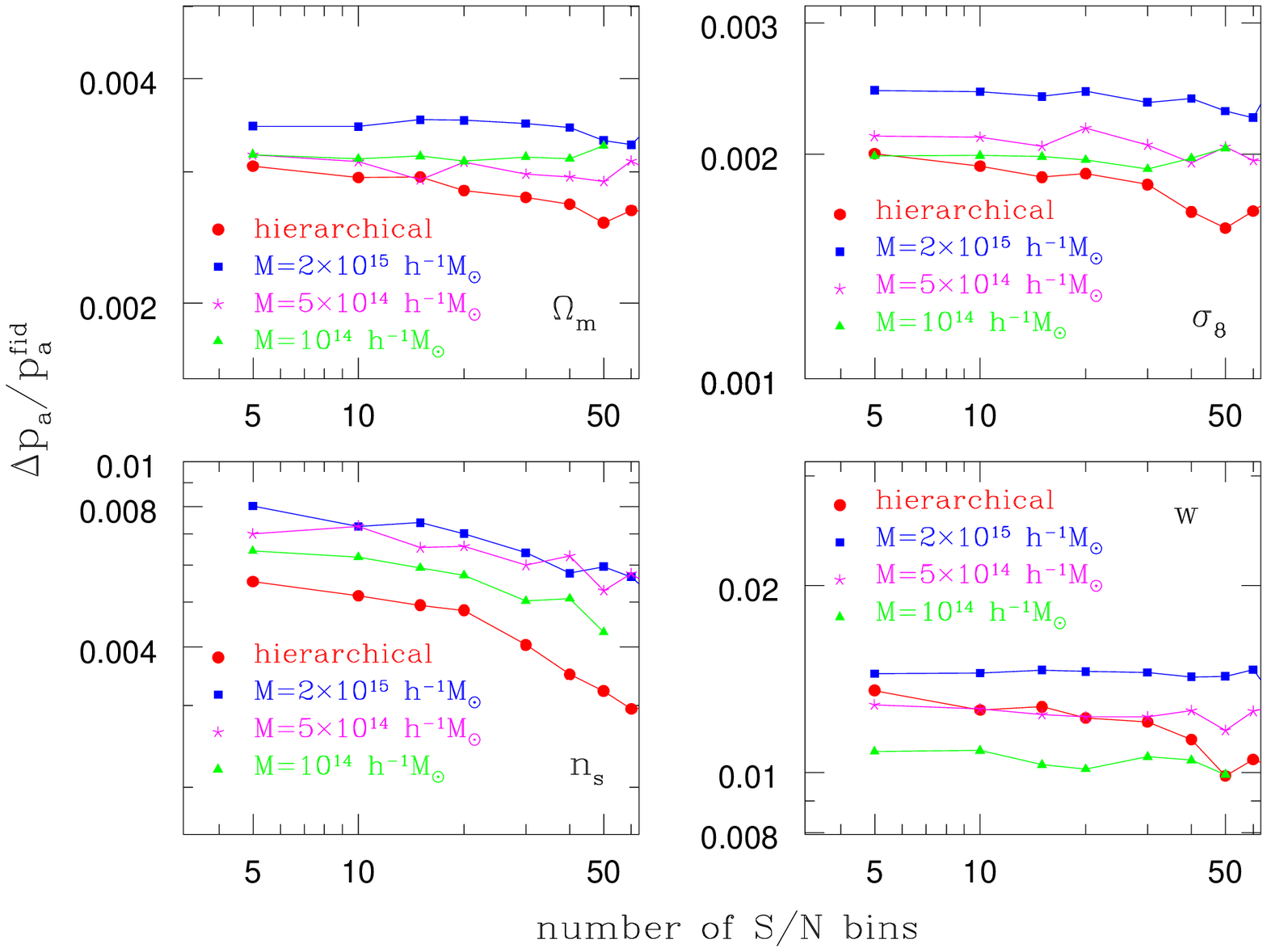}
\caption {Dependence of the fractional unmarginalized errors on the
  number of $\SN$ bins. We used \eqn{eq:Cinv_corrected} to estimate
  the Fisher matrix \eqn{eq:fisher_counts}. The hierarchical method
  performs best for a choice of $N_{\rm bin}=20$, while the
  single-filtering results seem largely independent of the number of
  bins used.}
\label{fig:SNbin_dependence}
\end{figure*}
We now address the issue of the number of bins used to estimate the
Fisher matrix in \eqn{eq:fisher_counts}. We expect that too coarse a
binning will diminish the constraints, as the information in the maps
would not be fully captured. We also expect the constraints to
saturate once a large enough number of bins is considered. However,
for all filtering methods we noticed a continuous improvement in the
constraints with the increasing number of bins. This is most likely
due to noise in the covariance matrix measurement.

As a remedy, we apply the correction discussed by
\citet{Hartlapetal2007}. Given a data set drawn from a multi-variate
Gaussian distribution, and given the maximum-likelihood estimator for
the covariance matrix C, i.e. \eqn{eq:cov2}, then an unbiased
estimator for the inverse covariance is:
\be
\widehat{\C ^{-1}}=\frac{N-N_{\rm bin}-2}{N-1}(\hat{\C})^{-1},\:  N_{\rm bin} < N-2,
\label{eq:Cinv_corrected}
\ee
where $N_{\rm bin}$ is the number of bins, and $N$ is the number of
realizations -- in our case the number of fiducial fields. We use
logarithmically-spaced $\SN$ bins in the interval $[2.8,
  14]$. Estimating our Fisher matrix with the above equation
alleviated significantly the dependence of the constraints on the
number of bins used, as seen in Figure \ref{fig:SNbin_dependence}.
The figure shows the dependence of the unmarginalized Fisher errors on
the number of bins in which the $\SN$ interval is divided. All errors
show little evolution with the number of bins, and are relatively
stable once $N_{\rm bin} \sim 20$, except for the case of $\ns$, where
there is slightly more evolution. The entire analysis presented in
this work was carried out for $N_{\rm bin}=20$.

\end{document}